\documentclass[onecolumn,a4paper,11pt,accepted=2024-01-05]{quantumarticle}
\pdfoutput=1

\usepackage{amsmath}
\usepackage{amsfonts}
\usepackage{amssymb}
\usepackage{mathtools}

\usepackage{uniinput}
\usepackage[utf8]{inputenc}
\usepackage[english]{babel}
\usepackage[T1]{fontenc}
\usepackage{xcolor}
\definecolor{riverlane_green}{RGB}{0, 111, 98}
\definecolor{riverlane_light_green}{RGB}{0, 150, 143}
\definecolor{riverlane_orange}{RGB}{255, 117, 0}
\definecolor{riverlane_red}{RGB}{220, 68, 5}
\definecolor{riverlane_pink}{RGB}{207, 111, 127}
\usepackage{hyperref}
\hypersetup{
  colorlinks   = true, 
  urlcolor     = riverlane_green, 
  linkcolor    = riverlane_orange, 
  citecolor   = riverlane_green  
}

\usepackage{braket}
\usepackage[compat=0.6]{yquant}
\useyquantlanguage{groups}
\usetikzlibrary{fit,quotes,matrix,positioning}
\usetikzlibrary{trees}

\pgfdeclareshape{yquant-multictrlshape}{
   \inheritsavedanchors[from=yquant-circle]%
   \foreach \anc in {center, north, north east, east, south east, south, south west, west, north west} {%
      \inheritanchor[from=yquant-circle]{\anc}%
   }%
   \inheritanchorborder[from=yquant-circle]%
   \backgroundpath{%
      \pgfpathmoveto{\pgfqpoint{-0.707107\dimexpr\xradius\relax}{-0.707107\dimexpr\yradius\relax}}%
      \pgfpathlineto{\pgfqpoint{0.707107\dimexpr\xradius\relax}{0.707107\dimexpr\yradius\relax}}%
      \pgfpathellipse{\pgfpointorigin}%
                     {\pgfqpoint{\xradius}{0pt}}%
                     {\pgfqpoint{0pt}{\yradius}}%
   }%
   \inheritclippath[from=yquant-circle]%
}

\yquantset{multictrl/.style={every control/.style={shape=yquant-multictrlshape,radius=1.05mm}}}

\yquantset{plusctrl/.style={/yquant/every control/.style={/yquant/operators/every not}, every positive control/.style={}}}

\usepackage{multirow}
\usepackage{amsthm}
\newtheorem*{theorem}{Theorem}
\newcommand{\sgn}{\operatorname{sgn}}

\usepackage[numbers,compress]{natbib}
\usepackage[nottoc,notlot,notlof]{tocbibind}

\begin{document}

\title{Block-encoding structured matrices for data input in quantum computing}
\author{Christoph Sünderhauf}
\affiliation{Riverlane, St.~Andrews House, 59 St.~Andrews Street, Cambridge CB2 3BZ, United Kingdom}
\email{christoph.sunderhauf@riverlane.com}
\author{Earl Campbell}
\affiliation{Riverlane, St.~Andrews House, 59 St.~Andrews Street, Cambridge CB2 3BZ, United Kingdom}
\affiliation{Dept.~of Physics and Astronomy, University of Sheffield, Sheffield S3 7RH, United Kingdom}
\author{Joan Camps}
\affiliation{Riverlane, St.~Andrews House, 59 St.~Andrews Street, Cambridge CB2 3BZ, United Kingdom}
\date{February 21st, 2023.~Accepted: January 5th, 2024}

\maketitle

\begin{abstract}
The cost of data input can dominate  the run-time of quantum algorithms. Here, we consider data input of arithmetically structured matrices via \emph{block encoding} circuits, the input model for the quantum singular value transform and related algorithms. We demonstrate how to construct block encoding circuits based on an arithmetic description of the sparsity and pattern of repeated values of a matrix. We present schemes yielding different subnormalisations of the block encoding; a comparison shows that the best choice depends on the specific matrix. The resulting circuits reduce flag qubit number according to sparsity, and data loading cost according to repeated values, leading to an exponential improvement for certain matrices.  We give examples of applying our block encoding schemes to a few families of matrices, including Toeplitz and tridiagonal matrices.
\end{abstract}

\tableofcontents

\section{Introduction}


The advent and astonishing increase in computational power of classical computing has truly revolutionised the world and ushered in the age of information. Yet, there are computational problems that are and will stay out of reach of classical computation due to their exponential complexity. Quantum computing \cite{nielsen_quantum_2010} offers the dazzling prospect to provide a speed up and move select problems from the intractable to the tractable. Demonstrations of first quantum computers \cite{arute_quantum_2019,noauthor_ibm_nodate,wu_strong_2021} up to a few hundred qubits provide a first step towards realising such a computational advantage. Apart from vast increases in the number and quality of qubits on the hardware level, improvements and development of new quantum algorithms are also dearly called for. Currently, the number of quantum algorithms and application use cases known to provide an exponential advantage are rather limited \cite{aaronson_how_2022,lee_is_2022}.

The inception of the quantum singular value transform (QSVT) \cite{gilyen_quantum_2019} has recently led to a new perspective on quantum algorithms. In what has been termed the ``grand unification of quantum algorithms'' \cite{martyn_grand_2021}, many previously known quantum algorithms have been reformulated within the framework of QSVT, including matrix inversion, phase estimation, Hamiltonian simulation, and amplitude amplification. The core of the QSVT algorithm is a polynomial transformation of an input matrix's singular values. Different choices of polynomials and input matrices give rise to these various applications.
In order to run a quantum algorithm in a real-world setting, data input (into the quantum computer) and data output (readout) are important steps and can severely limit any speed up provided by the quantum algorithm itself \cite{aaronson_read_2015}. In this article, we will study how to input structured data efficiently and provide a scheme that facilitates the construction of explicit quantum circuits for input of structured data matrices, demonstrated by several examples.

In QSVT based algorithms, the input model for matrices of data is that of a \emph{block encoding} \cite{gilyen_quantum_2019}.
Generally, an input matrix of data $A$ could be non-unitary and there is no quantum circuit that could implement the operator $A$ directly. Instead, in a block encoding, $A$ is embedded as a block inside a larger unitary $U$:
\begin{equation}
U = \begin{pmatrix}
A/α & \text{junk}_{i} \\
\text{junk}_{ii} & \text{junk}_{iii} \\
\end{pmatrix}
\label{eq:block encoding with junk}.
\end{equation}
Inside the larger unitary, the matrix $A$ is scaled down by the subnormalisation $α$. The precise values in the junk blocks are inconsequential (but must be consistent throughout one QSVT circuit). Subnormalisation and junk serve two purposes: First of all, they can be necessary to ensure that an embedding of $A$ into a unitary exists. However, the existence of an embedding (or even numerical knowledge of possible junk values) is not sufficient to input the data matrix $A$. Rather, $U$ must be implemented as a quantum circuit and expressed in an elementary quantum gate set in order to run it on a quantum computer. Finding such a quantum circuit will typically further increase the subnormalisation $α$ and the dimension of the junk blocks.
In a quantum circuit, the block encoding can be written as follows:
\begin{equation}
\begin{tikzpicture}
\begin{yquant}
qubit {flag qubits: $\ket{0}$} flags;
qubit {$\ket{j}$} block;
slash flags;
["north:$N$" {font=\protect\footnotesize, inner sep=0pt}]
slash block;
box {$U$} (-);
text {$\ket{0}$} flags;
text {$\ket{i}$} block;
discard flags;
discard block;
text {$= \ \ A_{ij}/α$.} (-);
\end{yquant}
\end{tikzpicture}
\end{equation}
The top qubit register consists of the flag qubits, its dimension depends on the dimensions of the junk blocks in \eqref{eq:block encoding with junk}. The block of the unitary containing $A/α$ is selected by initialising the flag qubits as $\ket{0}$ and postselecting them as $\ket{0}$. The probability of measuring the correct $\ket{0}$ outcome on all flag qubits is related to the subnormalisation; a smaller subnormalisation $α$ is better. A smaller subnormalisation typically also reduces the circuit length of a QSVT transforming the matrix, as lower resolution and lower polynomial degree are required. The bottom register has the same dimension $N$ as the matrix $A$. The matrix elements $A_{ij}/α$ can then be recovered as the amplitudes on the bottom register, as indicated in the circuit diagram.

Since block encodings are fundamental to QSVT, they have been studied previously. Quantum circuits implementing block encodings for arbitrary dense matrices have been worked out \cite{clader_quantum_2022,chakraborty_power_2019} but are exponentially expensive: For a 
$2^n\times2^n$ matrix on $n$ qubits, they require $O(2^n)$ $T$ gates.\footnote{The authors propose using a qRAM \cite{giovannetti_quantum_2008,hann_resilience_2021} to arrange the exponential number of $T$ gates in a linear \emph{depth} circuit, which however leads to an unpractical large exponential number of ancilla qubits.} 
However, the scope of these works did not include optimisations for matrices that are sparse or structured (have repeated values). Other work shows quantum circuit constructions for sparse matrices based on black-box oracles \cite{gilyen_quantum_2019}. There, the implementation of the oracles is not discussed, and complexity of the block encoding circuits is analysed in terms of black box usages. More specialised schemes for, eg.~density operators, POVM operators, Gram matrices \cite{gilyen_quantum_2019}, or kernel matrices \cite{nguyen_block-encoding_2022} have also been discussed. Explicit circuits for certain sparse, structured matrices are shown in \cite{camps_explicit_2022} for a specific value of $N$.

To gain a computational advantage for some problems, the exponential cost for arbitrary matrices \cite{clader_quantum_2022,chakraborty_power_2019} must be reduced. In this work, we consider matrices that are sparse and/or structured  (having repeated elements); with sufficient structure, we can construct exponentially more efficient circuits. We provide several variations of a block-encoding scheme based on oracles, each with a different subnormalisation. It depends on the matrix which variation performs best. The schemes fully take into account sparsity, reducing the number of flag qubits beyond schemes in \cite{gilyen_quantum_2019,clader_quantum_2022,chakraborty_power_2019}. Moreover, we explain how to construct circuit implementations of the required oracles given a family of matrix structures for increasing $N$, and provide several explicit examples.

Block encodings also appear beyond QSVT. In fact, there is a calculus allowing block-encoded matrices to be summed and multiplied \cite{gilyen_quantum_2019}.
Many modern quantum algorithms for chemistry are based on phase estimation via qubitisation \cite{low_hamiltonian_qubitisation_2019, babbush_encoding_2018, berry_qubitization_2019, lee_even_2021,ivanov_quantum_2022}, an application of quantum walks \cite{szegedy_quantum_2004,berry_hamiltonian_2015}. They are based on block-encoding the Hamiltonian. Successive articles go into great detail on how to construct the block encoding, and advances mainly stem from lower subnormalisation and shorter circuits of the block encoding. For chemistry applications, the block encodings are constructed as a linear combination of unitaries (LCU) with so-called PREPARE and UNPREPARE oracles. 
This viewpoint is quite distinct from other block-encoding schemes in the literature mentioned in the previous paragraphs; in section~\ref{sec:prep unprep block encoding}, we connect these two viewpoints, which can lead to an improved subnormalisation.

The quantum circuits involved are too long to directly run with noisy qubits provided by current hardware devices. Despite efforts to run QSVT primitives on noisy qubits \cite{Kikuchi_2023}, error correction and fault-tolerant quantum computation remains essential for longer circuits. In quantum error correction, an error correcting code is applied to several noisy physical qubits, yielding one or more logical qubits \cite{shor_scheme_1995}. The circuit is run at the level of logical qubits. Not all gates are equal in error correction codes, rather, in the popular surface codes \cite{fowler_surface_2012}, so-called $T$ gates and Toffoli gates (which can be reexpressed with $T$ gates) are much more expensive than Clifford gates \cite{bravyi_universal_2004,ogorman_quantum_2017,campbell_roads_2017,fowler_low_2019,blunt_perspective_2022}. Consequently, we will focus on $T$ gate count in assessing the cost of the block encoding.

\begin{table}[h]
    \centering
    \begin{tabular}{l@{\quad\quad}l@{\quad\quad}l@{\quad\quad}l@{}}
    \multicolumn{1}{@{}l}{Structured matrices (this work)} & \em{Data loading cost} & \em{Subnormalisation} &\em{Flag qubits} \\\hline
    Base (sec.~\ref{sec:base block encoding})& $D$ & $\sqrt{S_cS_r}||A||_\text{max}$ & $2 + \log_2 S$ \\
    Preamplified (sec.~\ref{sec:preamplified})& $D\cdot\text{amp}$ & $\sqrt{2}μ_p(A)$ & $5 + \log_2 S$ \\
    PREP (for $D\le S$) (sec.~\ref{sec:prep unprep block encoding})& $2D$ & $\sqrt{\sum_d |A_d|^{2p}\sum_d |A_d|^{2-2p}}$ & $1 + \log_2 S$\\
    \quad \rotatebox[origin=c]{180}{$\Lsh$} with optimal $p=1/2$ & $D$ & $\frac{\sqrt{S_cS_r}}{D}\sum_d |A_d|$ & $1 + \log_2 S$ \\[1.6ex]
    
    \multicolumn{4}{@{}l}{Sparse access matrices (Gily\'{e}n et.~al.~\cite{gilyen_quantum_2019})} \\\hline
    Base (\cite{gilyen_quantum_2019}, Lemma 48) & blackbox & $\sqrt{S_cS_r}||A||_\text{max}$ & $3 + \log_2 N$ \\
    Preamplified (\cite{gilyen_quantum_2019}, Lemma 49)& $\text{blackbox}\cdot \text{amp} $ & $\sqrt{2}μ_p(A)$ & $8+\log_2 N$\\[1.6ex]
    
    \multicolumn{4}{@{}l}{Explicit matrices (D.~Camps et.~al.~\cite{camps_explicit_2022})} \\\hline
    Banded circulant (\cite{camps_explicit_2022}, sec.~4.1) & $D=S$ & $S||A||_\text{max}$ & $1 + \log_2 S$ \\[1.6ex]
    
    \multicolumn{4}{@{}l}{Quantum data structure (Chakraborty et.~al.~\cite{chakraborty_power_2019},~Clader~et.~al.~\cite{clader_quantum_2022})} \\\hline
    Base (\cite{chakraborty_power_2019}, Lemma 6.2) & $N^2+N$ & $||A||_F$ & $1+\log_2 N$ \\
    $p$-norm (\cite{chakraborty_power_2019}, Lemma 6.1) & $2N^2$ & $μ_p(A)$& $2 + \log_2 N$ \\
    \end{tabular}
    \vspace{1ex}
    \begin{gather*}
        μ_p(A) = \sqrt{\max_i \sum_j|A_{ij}|^{2p}\max_j\sum_i|A_{ij}|^{2-2p}} \qquad
        \text{amp} \approx 3\left(\frac{γ_c}{δ}\log\frac{γ_c}{ε}+ \frac{γ_{r}}{δ}\log\frac{γ_{r}}{ε}\right) \\
        γ_c = \sqrt{\frac{S_c||A||_\text{max}^{2p}}{\sqrt{2}\max_i\sum_j |A_{ij}|^{2p}}}
        \qquad
        γ_r = \sqrt{\frac{S_r||A||_\text{max}^{2-2p}}{\sqrt{2}\max_j\sum_i |A_{ij}|^{2-2p}}}
    \end{gather*}
    \caption{Comparison of block-encoding schemes for a matrix $A$.
    $N$ dimension of the matrix. $D$ number of distinct data values $A_d$. $M$ maximum multiplicity of each value. $S_c, S_r$ maximum column and row sparsities (no.~non-zero elements per column/row). The maximum sparsity $S=\max(S_c,S_r)$ or multiplicity $M$ are padded to ensure $SN = DM$, if necessary.
    In the comparison we've added further factors to Gily\'{e}n et.~al.'s results to remove the $|A_{ij}|\le 1$ assumption.
    Block encoding from a quantum data structure is also discussed in \cite{clader_quantum_2022}.
    Lower is better for all of data loading cost, subnormalisation, and flag qubit number. $D$ data loading incurs $O(D)$ T-gate count, or $O(\sqrt{D})$ if a large number $O(\sqrt{D})$ of possibly dirty ancillas are available (see appendix~\ref{sec:data loading}).
    }
    \label{table:summary}
\end{table}

Section~\ref{sec:block encodings} explains and compares the proposed block encoding schemes. Their costs are summarised in Table~\ref{table:summary}, and an adaptation to yield Hermitian block encodings for symmetric matrices is considered in section~\ref{sec:hermitian}. Examples of using the scheme to construct block encodings of specific matrix families are provided in section \ref{sec:examples}, for example, our Toeplitz matrix encoding provides an exponential advantage to arbitrary matrix encoding. Finally, we draw some conclusions and give an outlook in section~\ref{sec:conclusions}. It is our hope that the schemes presented in this work will in the future enable the construction of efficient block encoding circuits for further families of matrices from a wide range of application areas. The appendices give more detail on the circuit implementations of some parts of the block encoding: Appendix~\ref{sec:data loading} focuses on the data loading oracles. Appendix~\ref{sec:singular value amplification} explains singular value amplification, a technique that can be employed to improve the subnormalisation, and provides some new analysis of its efficacy.

\section{Block-encoding schemes}
\label{sec:block encodings}

Throughout, we consider real matrices, including negative values. The schemes presented are adapted to structured matrices; that is, possibly sparse matrices with repeated data values and arithmetic descriptions of the structure available. This will become clear when constructing the circuits.

We take the following product as a total figure of merit for a block encoding:
\begin{equation}
\text{($T$-gate count)}\cdot\text{subnormalisation}.
\label{eq:figure of merit T-gate}\end{equation}
A lower subnormalisation increases the probability to measure $\ket{0}$ for the flag qubits, making it easier to extract the matrix. Lower subnormalisation typically also leads to shorter circuits in QSVT \cite{gilyen_quantum_2019, martyn_grand_2021} or qubitisation \cite{babbush_encoding_2018} algorithms that use the block encoding.
Taking the product \eqref{eq:figure of merit T-gate}
as the figure of merit is motivated as it is approximately constant under singular value amplification (see appendix~\ref{sec:singular value amplification})---which can reduce the subnormalisation of a block encoding by inversely increasing its circuit length (up to a logarithmic factor).
For simplicity, instead of considering $T$-gate count as a measure for circuit cost as in \eqref{eq:figure of merit T-gate}, we focus on ``data loading cost'', i.e.~the number of data values loaded, which generically dominates the cost. As we will see, we expect the other circuit parts to be implementable with $O(\text{polylog}\ N)$ $T$ gates. Of course, for low data loading cost, these other circuit parts could become dominant. Therefore we focus on the figure of merit
\begin{equation}
\text{(data loading cost)}\cdot\text{subnormalisation}.
\label{eq:figure of merit}
\end{equation}

The next subsections first describe our base block encoding scheme (section~\ref{sec:base block encoding}), followed by variants termed preamplified (section~\ref{sec:preamplified}) and PREP/UNPREP (section~\ref{sec:prep unprep block encoding}) that for some applications will improve the figure of merit \eqref{eq:figure of merit}.

\subsection{Base scheme}
\label{sec:base block encoding}

Let us introduce notation. Let $N$ be the dimension of the matrix $A$, and $D$ be the number of distinct data items $A_d$ ($0\le d<D$) in the matrix (apart from zeros following the sparsity pattern). Crucially, each data value may appear multiple times in the matrix. The multiplicity $M$ is the maximum multiplicity of any of the $D$ data items. Finally, we have column $S_c$ and row $S_r$ sparsities, the maximal number of non-zero elements per column or row (according to the sparsity pattern, if any) and maximum sparsity $S=\max(S_c, S_r)$. The smaller $S$, the more sparse the matrix.

\subsubsection{Simplified introductory case}
\label{sec:simplified}
Consider first the simple case in which each of the $D$ data items has the same multiplicity $M$, and each of the $N$ rows and columns the same sparsity $S_c=S_r=S$, and all of these are powers of 2. The equality
\begin{equation}
    MD = NS_c = NS_r = \#\text{nonzero} \label{eq:nonzero}
\end{equation}
then follows from calculating the number of nonzero entries in the matrix's sparsity pattern from three different perspectives. From these three perspectives, each nonzero matrix element is labelled either by
\begin{itemize}
\item $(d,m)$, its data index $0\le d<D$ and multiplicity $0\le m<M$, or
\item $(j,s_c)$, its column index $0\le j<N$ and column sparsity index $0\le s_c<S_c$, or
\item $(i,s_r)$, its row index $0\le i<N$ and row sparsity index $0\le s_r<S_r$.
\end{itemize}
The oracles underlying our block encoding are the unitary column $O_c$ and row $O_r$ oracles relating these three equivalent descriptions. In particular,
\begin{equation}
    O_c\ket{d}\ket{m} = \ket{j}\ket{s_c},\ O_r\ket{d}\ket{m} = \ket{i}\ket{s_r}.
\end{equation}

We suggest to construct these oracles by first establishing a labelling of the matrix by $(d,m)$. We suppose that arithmetic expressions for $i(d,m)$ and $j(d,m)$ follow; throughout we use the term ``structured matrix'' to refer to this kind of arithmetic structure allowing the computation of the positions of elements. These arithmetic expressions can then be converted to quantum circuits, possibly making use of ancillas. The exact values of $s_r$ and $s_c$ are irrelevant as long as they fall within range; they arise naturally in the conversion of the arithmetic expressions, making the quantum circuits unitary.
Example constructions are performed in section~\ref{sec:examples}. Quantum arithmetics can be implemented efficiently \cite{gidney_halving_2018,sanders_compilation_2020}, and so we expect the same of $O_c$ and $O_r$. Provided the arithmetic expressions underlying the oracles are sufficiently short (for example, do not increase in length with $N$), the $T$-gate counts of the oracles $O_c$ and $O_r$ are expected to be of order $O(\text{polylog}\ N)$ \cite{gidney_halving_2018,sanders_compilation_2020}. 

In addition to the structure of the matrix, which is encoded in the oracles $O_c$ and $O_r$, the oracle 
\begin{equation}
    O_\text{data} = \sum_{d=0}^{D-1}
    R_X(2\arccos A_d / ||A||_\text{max})\otimes\ket{d}\!\bra{d}
    \label{eq:formula multiplexed rotations}
\end{equation}
encoding the values of the $D$ data items is needed. The factor $||A||_\text{max} = \max_d |A_d|$ is needed to ensure all values are in-range of the arccosine. 
Provided the first qubit is initialised and postselected as $\ket{0}$, the effect of $O_\text{data}$ is loading the correct values:
\begin{equation}
O_\text{data}\ket{0}\otimes\ket{d} = \left(A_d/||A||_\text{max}\ket{0} -i \sqrt{1-(|A_d|/||A||_\text{max})^2}\ket{1}\right)\otimes\ket{d}. 
\end{equation}
In a quantum circuit, we write these multiplexed rotations using one flag qubit as
\begin{equation}
\begin{tikzpicture}
\begin{yquant}
qubit {data} a;
qubit {$\ket{d}$} d;
["north east:$D$" {font=\protect\footnotesize, inner sep=0pt}]
slash d;
[multictrl]
box {$R(A_d/||A||_\text{max})$} a ~ d;
\end{yquant}
\end{tikzpicture}.
\label{eq:circuit multiplexed rotations}
\end{equation}
The notation of a slash in a control indicates that the gate is not controlled on a single value, but multiplexed for various values of the control register as in \cite{low_trading_2018,clader_quantum_2022}.

These multiplexed rotations are the data loading step and may be implemented in multiple ways. See appendix~\ref{sec:data loading} for details. In general, the data loading cost of \eqref{eq:circuit multiplexed rotations} corresponds to a Toffoli count of $O(D)$ (QROM \cite{babbush_encoding_2018}) or $O(\sqrt{D})$ (select/swap network \cite{low_trading_2018}) for the multiplexing, plus $T$ gates to synthesize rotations. Note there is no dependence on $M,N$ or $S$. As with the other oracles, implementation details possibly require ancilla qubits.

The oracles are supplemented by $H_S$ to construct the block encoding.
The gate $H_S$, sometimes called diffusion operator, creates an equal superposition state
\begin{equation}
    H_S\ket{0} = \frac{1}{\sqrt{S}}\sum_{s=0}^{S-1} \ket{s}.
\end{equation}
If $S$ is a power of 2, $H_S$ is simply a string of Hadamard gates. In other cases, it can for example be constructed by amplitude amplification that uses an ancilla (see \cite{babbush_encoding_2018}).
Putting these together, the block encoding is
\begin{equation}
\begin{tikzpicture}
\begin{yquant}
    qubit {data} f0;
    qubit {s} f2;
    qubit {$\ket{j}$} block;
    
    ["north east:$S$" {font=\protect\footnotesize, inner sep=0pt}]
    slash f2;
    ["north east:$N$" {font=\protect\footnotesize, inner sep=0pt}]
    slash block;
    hspace {5pt} -;
    box {$H_S$} f2;
    
    align -;
    text {$s_c$} f2;
    text {$j$} block;
    box {$O_c^\dag$} (f2, block);
    setstyle {thick, riverlane_pink} f2, block;
    text {$d$} f2;
    text {$m$} block;
    hspace {-3pt} -;
    ["north east:$D$" {font=\protect\footnotesize, inner sep=0pt}]
    slash f2;
    ["north east:$M$" {font=\protect\footnotesize, inner sep=0pt}]
    slash block;
    hspace {5pt} -;
    
    [multictrl]
    box {$R(A_d/||A||_\text{max})$} f0 ~ f2;

    align -;
    text {$d$} f2;
    text {$m$} block;
    box {$O_r$} (f2, block);
    setstyle {black} f2, block;
    text {$s_r$} f2;
    text {$i$} block;
    
    box {$H_S^\dag$} f2;
    output {$\ket{i}$} block;
\end{yquant}
\end{tikzpicture}.
\label{eq:simplified block encoding}
\end{equation}
Crucially, the pink qubit registers in the middle can have dimensions $D$ and $M$ distinct from the black qubit registers of dimension $S$ and $N$. Yet, thanks to the equality $NS=MD$, the oracles have same total input and output dimensions. To recover the matrix from the block encoding, the flag qubits (i.e.~the data qubit and the s register) must be initialised and postselected as $\ket{0}$. The values
$\frac{A_{ij}}{S||A||_\text{max}}$
are then recovered when initialising the bottom register with $\ket{j}$ and postselecting/measuring an $\ket{i}$. This can be most easily seen by inserting a resolution of the identity $\mathbb{I} = \sum_d\sum_m \ket{d}\!\bra{d}\otimes \ket{m}\!\bra{m}$ into the middle of the circuit:
\begin{align}
& \bra{0}^\text{data}\bra{0}^\text{s}\bra{i} H_S^\dagger O_r O_\text{data} \left(\sum_{d=0}^{D-1}\sum_{m=0}^{M-1} \ket{d}\!\bra{d}\otimes \ket{m}\!\bra{m}\right) O_c^\dag H_S \ket{0}^\text{data}\ket{0}^\text{s}\ket{j} \nonumber\\
    &{} =  \sum_{d=0}^{D-1}\sum_{m=0}^{M-1} \frac{A_d}{||A||_\text{max}} \bra{0}^\text{s}\!\bra{i}H_S^\dag O_r\ket{d}\!\ket{m}\bra{d}\!\bra{m}O_c^\dag H_S\ket{0}^\text{s}\!\ket{j} \nonumber\\
    &{} = \sum_{d=0}^{D-1}\sum_{m=0}^{M-1} \frac{A_d}{||A||_\text{max}} \frac{1}{\sqrt{S}} δ_{i = i(d,m)} \frac{1}{\sqrt{S}} δ_{j =j(d,m)}  =\frac{A_{ij}}{S||A||_\text{max}}.
\end{align}
The factor of $S$ comes from the $H_S$ gates. Thus, the circuit is a block encoding of $A$ with subnormalisation $S||A||_\text{max}$ and $1 + \log_2 S$ flag qubits of total dimension $2S$. The matrix is implemented exactly, apart from any finite accuracy in the multiplexed rotations (the data loading).

In applications of the block encoding scheme, we expect the matrix structure (the sparsity pattern and pattern of repeated values) to be given, from which $O_c$ and $O_r$ can be inferred. Different instances of the problem only require replacement of the data loading oracle, a straightforward automatable task.

\subsubsection{Full block encoding}
A general structured matrix may not fulfill the strict requirements of the simplified case in the preceding section~\ref{sec:simplified}. Here, we consider the general case, where each of the $D$ data items may have a distinct multiplicity (with maximum $M$), and each of the $N$ rows and columns may have distinct sparsities (with maximum $S_c$ for columns and $S_r$ for rows, and $S=\max(S_c, S_r)$). Therefore, a priori, $MD=NS$ may not hold. We pad $M$ and/or $S$, increasing their size with further dummy index range, until
\begin{equation}
    MD=NS \label{eq:padded equality}
\end{equation}
is fulfilled and the same construction with oracles $O_c$ and $O_r$ is possible. The action of these oracles on the padded dummy indices is insignificant (they will be flagged and deleted by an out-of-range oracle introduced in the next paragraph), as long as unitarity is ensured. Without loss of generality, we assume the resulting qubit register dimensions are powers of 2; otherwise they can trivially be embedded in a larger register made up of an integer number of qubits.

As in the simplified case, the matrix is labelled from the three perspectives $(d,m)$, $(i,s_r)$, and $(j,s_c)$. Now however, because of the padding, not all labels $(d,m)$ with $0\le d<D, 0\le m<M$ are in-range and are mapped by $O_r$ and $O_c$ to row and column indices $i,j$ matching the matrix pattern. These are flagged by a new oracle, which we call $O_{\text{rg}}$, the out-of-range oracle:
\begin{equation}
O_\text{rg} \ket{d}\ket{m}\ket{0} = \begin{cases}
\ket{d}\ket{m}\ket{0}\ &\text{if}\ A_{i(d,m),j(d,m)} =A_d \\
\ket{d}\ket{m}\ket{1}\ &\text{if}\ A_{i(d,m),j(d,m)} =0.
\end{cases}
\end{equation}
The oracle is controlled on the $D,M$ registers and flips a ``delete" flag qubit whenever $(d,m)$ is out-of-range, which we draw as
\begin{equation}
\begin{tikzpicture}
\begin{yquant}
qubit del;
qubit {$d$} d;
qubit {$m$} m;

setstyle {thick, riverlane_pink} d,m;

["north east:$D$" {font=\protect\footnotesize, inner sep=0pt}]
slash d;
["north east:$M$" {font=\protect\footnotesize, inner sep=0pt}]
slash m;
hspace {7pt} -;

[plusctrl, shape=yquant-circle]
box {$O_{\text{rg}}$} (d,m) | del;
\end{yquant}
\end{tikzpicture}.
\end{equation}
Reminiscent of the usual circular notation for controls, the circle/oval serves as the control for the $O_\text{rg}$-controlled NOT; the $d$ and $m$ registers are not modified. The out-of-range oracle can be implemented with quantum arithmetics, possibly using ancilla qubits, and have a low $O(\text{polylog}\ N)$ $T$-gate count. Like for the $O_c$ and $O_r$ oracles, the arithmetic expressions underlying $O_\text{rg}$ follow from the structure of the matrix and the chosen labelling in terms of $(d,m)$.

The oracles $O_c$ and $O_\text{data}$ may map invalid $(d,m)$ to any values, they just need to be unitary. For valid $(d,m)$, the specific values of the corresponding $s_r$ and $s_c$ are irrelevant, as long as they are within range $0\le s_r<S_r$ and $0\le s_c<S_c$. Example constructions are shown in section~\ref{sec:examples}.

The full block encoding circuit is:
\begin{equation}
\begin{tikzpicture}
\begin{yquant}
    qubit {data} f0;
    qubit del;
    qubit {s} f2;
    qubit {$\ket{j}$} block;
    
    ["north east:$S$" {font=\protect\footnotesize, inner sep=0pt}]
    slash f2;
    ["north east:$N$" {font=\protect\footnotesize, inner sep=0pt}]
    slash block;
    hspace {5pt} -;
    box {$H_{S_c}$} f2;
    
    align -;
    text {$s_c$} f2;
    text {$j$} block;
    box {$O_c^\dag$} (f2, block);
    setstyle {thick, riverlane_pink} f2, block;
    text {$d$} f2;
    text {$m$} block;
    hspace {-3pt} -;
    ["north east:$D$" {font=\protect\footnotesize, inner sep=0pt}]
    slash f2;
    ["north east:$M$" {font=\protect\footnotesize, inner sep=0pt}]
    slash block;
    hspace {5pt} -;

    [plusctrl, shape=yquant-circle]
    box {$O_\text{rg}$} (block, f2) | del;

    [multictrl]
    box {$R(A_d/||A||_\text{max})$} f0 ~ f2;

    align -;
    text {$d$} f2;
    text {$m$} block;
    box {$O_r$} (f2, block);
    setstyle {black} f2, block;
    text {$s_r$} f2;
    text {$i$} block;
    
    box {$H_{S_r}^\dag$} f2;
    output {$\ket{i}$} block;
\end{yquant}
\end{tikzpicture}.
\label{eq:block encoding circuit}
\end{equation}
It has one flag qubit (the del qubit) more than the simplified scheme, giving $2 + \log_2 S$ flag qubits.
The subnormalisation now takes into account possibly unequal $S_c\neq S_r\neq S$ and is
\begin{equation}
    \sqrt{S_cS_r}||A||_\text{max}.
\end{equation}
The data loading oracle must load $D$ data items. Because the other oracles can be implemented with arithmetics in $O(\text{polylog}\ N)$ $T$-count \cite{gidney_halving_2018,sanders_compilation_2020}, we take the number of data items to load as a metric for the cost of the block encoding circuit. Note that if $D$ is small and constant as the size $N$ grows, eventually the cost of the other oracles will surpass the data loading.

The cost figures (data loading cost, subnormalisation, flag qubit dimension) are summarised in table~\ref{table:summary}. As a reference, we also include the cost figures from encoding schemes found in the literature. Gily\'{e}n et.~al.~present a block encoding scheme for sparse access matrices (\cite{gilyen_quantum_2019} Lemma 48). In contrast to our scheme, where the data loading oracle is multiplexed only on the $D$ distinct indices, in their scheme the data loading oracle is multiplexed on the $N$ row and $N$ column indices $i$ and $j$. The inner workings of the oracle are not specified. See appendix~\ref{sec:sparse_data_loading} for a possible efficient implementation. The subnormalisation is the same as in our case. While they have $3+\log_2 N$ flag qubits, our encoding takes into account the sparsity and requires only $2 + \log_2 S$ flag qubits.

Chakraborty et.~al.~present a block encoding scheme for a matrix from a quantum data structure (\cite{chakraborty_power_2019}, Lemma 6.2). A similar construction is also discussed by Clader et.~al.~\cite{clader_quantum_2022}. Regardless of the sparsity, the data loading cost is always $N^2+N$ items. The works presuppose that this data loading can be performed efficiently with a quantum data structure like qRAM \cite{giovannetti_quantum_2008,hann_resilience_2021}. It promises a logarithmic $T$-gate depth. However, this is achieved by a large parallel execution of $T$ gates on a large number of ancillary qubits. The total number of $T$ gates is not reduced compared to other approaches (QROM \cite{babbush_encoding_2018}, select-swap \cite{low_trading_2018}). In the comparison table~\ref{table:summary} we remain agnostic about the type of data loading (qRAM, QROM,…) and record the number of items to be loaded.
When $D\le N$, we find a lower figure of merit~\eqref{eq:figure of merit}, $\text{(data loading cost)}\cdot\text{subnormalisation}$, for our block encoding scheme:
\begin{equation}
(N^2+N)||A||_F \ge  N^2 ||A||_\text{max} \ge D \sqrt{S_cS_r} ||A||_\text{max}.
\end{equation}
However, for a matrix with more distinct data entries than $N$, the scheme by Chakraborty may or may not be advantageous to the base scheme presented above, depending on the matrix norms.

D.~Camps et al.~\cite{camps_explicit_2022} show an explicit example construction for a circulant matrix. Similarly to our scheme, it requires only $D$ data loading even though values appear multiple times, and a flag qubit number of $\approx\log_2 S$. However, the scheme is only suitable for matrices with a very particular structure: Each value must appear exactly once in each column. Our block-encoding encompasses this case and goes beyond, covering other types of structured matrices.

\subsection{Preamplified scheme}
\label{sec:preamplified}
In this section, we will show how the block encoding from the base scheme can, in some cases, be improved by a method called preamplification.
The subnormalisation of a block encoding can be reduced by performing singular value amplification (see app.~\ref{sec:singular value amplification}) \cite{gilyen_quantum_2019,chakraborty_power_2019}, resulting in an $ε$-approximate block encoding with subnormalisation reduced by an amplification factor $γ$. However, it requires $O\left(\frac{γ}{δ}\log\frac{γ}{ε}\right)$ applications of the original block encoding ($δ$ is related to a bound on the matrix's singular values). Simply performing singular value amplification on the full block encoding therefore does not improve the figure of merit \eqref{eq:figure of merit}, (data loading cost)$\cdot$subnormalisation; in fact, because of the factors related to the accuracy of the approximate result, it gets worse. 
Preamplification, presented by Gily\'{e}n et.~al.~\cite{gilyen_quantum_2019}, is based on two individual singular value amplifications of two separate circuit parts. As we will see, it is also applicable to our block encoding and can improve its figure of merit \eqref{eq:figure of merit} in some cases, intuitively when the matrix has values of strongly varying magnitude.

The starting point for preamplification is to split the data loading oracle into two parts according to a choice of $0\le p \le 1$. The block encoding circuit \eqref{eq:block encoding circuit} becomes:
\begin{equation}
\begin{tikzpicture}[scale=0.9]
\begin{yquant}
    qubit {data0} f0a;
    qubit {data1}f0b;
    qubit {del} f1;
    qubit {s} f2;
    qubit {$\ket{j}$} block;
    
    ["north east:$S$" {font=\protect\footnotesize, inner sep=0pt}]
    slash f2;
    ["north east:$N$" {font=\protect\footnotesize, inner sep=0pt}]
    slash block;
    hspace {8pt} -;
    
    [name=leftD]
    box {$H_{S_c}$} f2;
    
    align -;
    text {$s_c$} f2;
    text {$j$} block;
    [name=Oc]
    box {$O_c^\dag$} (f2, block);
    setstyle {thick, riverlane_pink} f2, block;
    text {$d$} f2;
    text {$m$} block;
    hspace {-3pt} -;
    ["north east:$D$" {font=\protect\footnotesize, inner sep=0pt}]
    slash f2;
    ["north east:$M$" {font=\protect\footnotesize, inner sep=0pt}]
    slash block;
    
    [multictrl, name=leftR]
    box {$R\left(\frac{\sgn(A_d)|A_d|^p}{\max_{d'} |A_{d'}|^p}\right)$} f0a ~ f2;
    
    hspace {5pt} -;
    [plusctrl, shape=yquant-circle]
    box {$O_\text{rg}$} (f2, block) | f1;
    hspace {5pt} -;

    [multictrl, name=rightR]
    box {$R\left(\frac{|A_d|^{1-p}}{\max_{d'} |A_{d'}|^{1-p}}\right)$} f0b ~ f2;
    
    align -;
    text {$d$} f2;
    text {$m$} block;
    [name=Or]
    box {$O_{r}$} (f2, block);
    setstyle {black} f2, block;
    text {$s_r$} f2;
    text {$i$} block;
    [name=rightD]
    box {$H_{S_r}^\dag$} f2;
    hspace {5pt} -;
    output {$\ket{i}$} block;
    
    \node[draw, dashed, fit=(leftD) (leftR) (Oc), "$U_c^\dag$" below] {};
    \node[draw, dashed, fit=(rightD) (rightR) (Or), "$U_r$" below] {};
\end{yquant}
\end{tikzpicture}
\label{eq:circuit preamplification split}
\end{equation}
The two data qubits and data loading oracles combine to give the same matrix as the original block encoding. In fact, also the subnormalisation
\begin{equation}
    \left(\sqrt{S_c}\max_{d} |A_d|^p\right)\left(\sqrt{S_r}\max_d |A_d|^{1-p} \right) = \sqrt{S_cS_r}\max_{d} |A_d| = \sqrt{S_cS_r}||A||_\text{max}
\end{equation}
has not changed.
The idea of preamplification is that, as we will see momentarily, the two circuit parts $U_c^\dag$ and $U_r$ are block-encodings in their own right, and can be singular value amplified individually by amplification factors $γ_c$ and $γ_r$. While the data loading cost increases by a factor of $O(γ_c+γ_r)$ (dropping logarithmic factors), the total subnormalisation of the matrix reduces multiplicatively by $γ_cγ_r$. Depending on the factors hidden in the big-$O$ notation, preamplification can thereby reduce the figure of merit (data loading cost)$\cdot$subnormalisation.

Let us discuss $U_c^\dag$, where $U_r$ is similar. The unitary $U_c^\dag$ is a slightly more general block encoding, where the flag qubits on the left in circuit \eqref{eq:circuit preamplification split} are data0 through s, and on the right only data0 through del. Omitting data1 and del that do not participate in $U_c^\dag$, the encoded non-square matrix can be deduced by inserting resolutions of identity $\sum_d \ket{d}\!\bra{d}, \sum_m \ket{m}\!\bra{m}, \sum_j \ket{j}\!\bra{j}$:
\begin{align}
    \bra{0_\text{data0}}U_c^\dag\ket{0_\text{data0}}\!\ket{0_\text{s}} &=
    \sum_{j=0}^{N-1}\sum_{d=0}^{D-1}\sum_{m=0}^{M-1} \ket{d}\!\ket{m} \bra{d}\!\bra{m} \frac{\sgn(A_d)|A_d|^p}{\max_{d'}|A_{d'}|^p}O_c^\dag  H_{S_c}\ket{0_\text{s}}\!\ket{j}\bra{j} \\
    &= \sum_{j=0}^{N-1}\sum_{d=0}^{D-1}\sum_{m=0}^{M-1}\sum_{s_c=0}^{S_c-1}\frac{1}{\sqrt{S_c}} \ket{d}\!\ket{m} \frac{\sgn(A_d)|A_d|^p}{\max_{d'}|A_{d'}|^p} \underbrace{\bra{d}\!\bra{m}O_c^\dag \ket{s_c}\!\ket{j}}_{\mathclap{\text{fixes $(d,m)$ corresponding to $(j,s_c)$}}} \bra{j} \\
    &= \sum_{j=0}^{N-1}\underbrace{\sum_{s_c=0}^{S_c-1} \ket{d}{\ket{m}}\frac{\sgn(A_d)|A_d|^p}{\sqrt{S_c}\max_{d'} |A_{d'}|^p}}_{\ket{x_j}}\bra{j}
\end{align}
with $d=d(j,s_c),m=m(j,s_c)$ according to $O_c^\dag$. This matrix is already written in the form of a singular value decomposition because $\{\ket{x_j}\}$ and $\{\ket{j}\}$ are orthogonal systems. The singular values follow as their normalisation
\begin{equation}
    \zeta_j = \sqrt{\braket{x_j|x_j}} = \sqrt{\frac{\sum_{i=0}^{N-1} |A_{ij}|^{2p}}{S_c\max_{d}|A_{d}|^{2p}}}.
    \label{eq:singular value to be amplified}
\end{equation}
Following \cite{gilyen_quantum_2019}, we choose the amplification factor $γ_c$ as follows (see appendix~\ref{sec:singular value amplification} for details):
\begin{equation}
    γ_c=\frac{1}{\sqrt[4]{2}}\frac{1}{\max_j \zeta_j} = \max_{d}|A_d|^p \sqrt{\frac{S_c}{\sqrt{2}\max_j \sum_i |A_{ij}|^{2p}}}.
    \label{eq:gamma_c}
\end{equation}

Replacing $U_c^\dag$ and $U_r$ by their amplification circuits results in a total subnormalisation divided by $γ_cγ_r$, such that the new subnormalisation after preamplification is
\begin{equation}
    \frac{\sqrt{S_c}\max_d |A_d|^p\cdot\sqrt{S_r}\max_d |A_d|^{1-p}}{γ_cγ_r}=\sqrt{2\max_j\sum_i |A_{ij}|^{2p}\max_i\sum_j |A_{ij}|^{2-2p}},
    \label{eq:subnormalisation preamplified}
\end{equation}
which we record in table~\ref{table:summary}. Two flag qubits are needed in addition to \eqref{eq:circuit preamplification split} to perform the two amplifications.
The data loading cost will be
\begin{equation}
    \approx D \cdot 3\left(\frac{γ_c}{δ} \log\frac{γ_c}{ε} + \frac{γ_r}{δ} \log\frac{γ_r}{ε}\right),
    \label{eq:preamplification data loading}
\end{equation}
see appendix~\ref{sec:singular value amplification} where we have determined the prefactor 3. The largest possible choice of $δ$ is $δ=1-1/\sqrt[4]{2}\approx0.16$, see \eqref{eq:min delta} in the appendix~\ref{sec:singular value amplification}. The accuracy $ε$ determines the accuracy of the amplifications of $U_c^\dag$ and $U_r$. Hence, the accuracy of the full matrix is bounded by $(2ε+ε^2)\cdot \text{subnormalisation}$. 

Preamplification was introduced in the scheme by Gily\'{e}n et.~al.~\cite{gilyen_quantum_2019}, where it works the same way. Like in the comparison of the base block encoding, our scheme requires a lower number of flag qubits for sparse matrices ($S<N$).

While the $p$-norm encoding from the quantum data structure scheme (\cite{chakraborty_power_2019}) does not employ singular value amplification, it is similar in spirit because it splits the matrix element into powers by $p$ and $1-p$. The subnormalisation is the same as in preamplification, except for the $\sqrt{2}$ factor. To compare the data loading cost, bound the amplification factors by $γ_c\le \sqrt{S_c/\sqrt{2}}, γ_r\le \sqrt{S_r/\sqrt{2}}$. Then the preamplified data loading cost (dropping logarithms) obeys
$D\cdot O(γ_c+γ_r) \le O(D\sqrt{S}),$
which is to be contrasted with $2N^2$ from $p$-norm encoding. If the matrix has $D=N^2$ different values, $p$-norm encoding is clearly favourable. Our scheme does well if the matrix is structured with a lower $D$.

Preamplification has a significant overhead in data loading cost, due to the prefactor, $δ$ and $ε$ in \eqref{eq:preamplification data loading}. In order to still achieve a benefit over the base block encoding scheme, the amplification factors $γ_c$ \eqref{eq:gamma_c} and $γ_r$ 
must be large.
Intuitively, this requires matrices with values of strongly varying magnitude.
Whether the base or preamplified schemes yield a better block encoding w.r.t.~$\text{(data loading cost)}\cdot\text{subnormalisation}$ depends on the specific matrix.
The optimal choice of $p$ in the preamplification scheme depends on the specific matrix. 

\subsection{PREP/UNPREP scheme}
\label{sec:prep unprep block encoding}
When the matrix structure has repeated data such that $D\le S_c,S_r$, often, the row/column oracle and the multiplexed rotations commute. Intuitively, this is the case when each data values appears in all (or most) of the rows and columns and the structure of the matrix is such that the columns and rows are (mostly) permutations of each other. Then, assuming commutativity, one can use a PREP/UNPREP scheme to reduce the subnormalisation of the base block encoding. PREP and UNPREP operators previously appeared in quantum algorithms for chemistry, when block-encoding Hamiltonians \cite{babbush_encoding_2018}. We show how such operators can be used in our block encoding scheme for more general structured matrices.

The starting point is splitting the multiplexed rotations into two parts \eqref{eq:circuit preamplification split}, just like in preamplification. By assumption, the row/column oracles commute with the rotations, hence we can use the following identity:
\begin{equation}
\begin{tikzpicture}[baseline=(current bounding box.center)]
\begin{yquantgroup}
\registers{
qubit {} a;
qubit {} b;
}
\circuit{
init {$\ket{0}$} a,b;
["north east:$D$" {font=\protect\footnotesize, inner sep=0pt}]
slash b;
hspace {5pt} -;
box {$H_D$} b;

[multictrl]
box {$R\left(\frac{\sgn(A_d) |A_d|^p}{\max_{d'} |A_{d'}|^p}\right)$} a ~ b;

output {$\ket{0}$} a;
}
\end{yquantgroup}
\end{tikzpicture}
\quad=\quad \underbrace{\left(\frac{\sqrt{\sum_d |A_d|^{2p}}}{\sqrt{D}\max_d |A_d|^{p}}\right)}_{{}:=X}\cdot\,\text{PREP}\ket{0}
\label{eq:prep identity}
\end{equation}
with a prepare operator acting as
\begin{equation}
\text{PREP}\ket{0} = \frac{1}{\sqrt{\sum_d |A_d|^{2p}}}\sum_d \sgn(A_d)|A_d|^p \ket{d}
\label{eq:prep}
\end{equation}
and, because the left side is not a normalised quantum state, a scaling factor $X$.
While $\text{PREP}\ket{0}$ is of course a normalised state, the left hand side of \eqref{eq:prep identity} is not, due to the postselection of the flag qubit as $\ket{0}$. The quotient factor $X$ in \eqref{eq:prep identity} is smaller than one and results in a lower subnormalisation after application of the identity.
A state preparation operator for a $D$-dimensional state can be implemented in various ways with no more than $D$ data loading cost, see appendix~\ref{sec:state preparation} for details.
We use a similar operator on the right side of the circuit,
\begin{equation}
\bra{0}\text{UNPREP} = \frac{1}{\sqrt{\sum_d|A_d|^{2-2p}}}\sum_d |A_d|^{1-p}\bra{d}.
\label{eq:unprep}
\end{equation}
The total block encoding circuit is 
\begin{equation}
\begin{tikzpicture}
\begin{yquant}
    qubit del;
    qubit d;
    qubit {s} f2;
    qubit {$\ket{j}$} block;

    setstyle {thick, riverlane_pink} d;
    
    ["north east:$D$" {font=\protect\footnotesize, inner sep=0pt}]
    slash d;
    ["north east:$\frac{S}{D}$" {font=\protect\footnotesize, inner sep=0pt}]
    slash f2;
    ["north east:$N$" {font=\protect\footnotesize, inner sep=0pt}]
    slash block;
    hspace {9pt} -;
    box {$H_{S_c/D}$} f2;
    box {PREP} d;
    
    align -;
    [shape=yquant-init, decoration={mirror}] 
    inspect {$s_c$} (d,f2);
    text {$j$} block;
    box {$O_c^\dag$} (f2, block, d);
    setstyle {thick, riverlane_pink} f2, block;
    discard f2;
    text {$d$} d;
    text {$m$} block;
    hspace {-3pt} -;
    ["north east:$D$" {font=\protect\footnotesize, inner sep=0pt}]
    slash d;
    ["north east:$M$" {font=\protect\footnotesize, inner sep=0pt}]
    slash block;
    hspace {5pt} -;

    [plusctrl, shape=yquant-circle]
    box {$O_\text{rg}$} (block, f2,d) | del;

    align -;
    text {$d$} d;
    text {$m$} block;
    box {$O_r$} (f2, block,d);
    settype {qubit} f2;
    setstyle {black} f2, block;
    inspect {$s_r$} (f2,d);
    text {$i$} block;
    box {UNPREP} d;
    box {$H_{S_r/D}^\dag$} f2;
    output {$\ket{i}$} block;
\end{yquant}
\end{tikzpicture}
\label{eq:circuit prep block encoding}
\end{equation}
with subnormalisation
\begin{equation}
  α_p = \sqrt{\frac{S_c}{D}} \sqrt{\sum_d |A_d|^{2p}} \sqrt{\frac{S_r}{D}}\sqrt{\sum_d |A_d|^{2-2p}} = \frac{\sqrt{S_cS_r}}{D}\sqrt{\sum_d |A_d|^{2p}\sum_d|A_d|^{2-2p}}.
\end{equation}
Contrary to preamplification, there is no singular value amplification that will necessitate repeated data loading. Instead, data will just need to be loaded twice (i.e. $2D$ data loading cost), for PREP and for UNPREP.

An application of Callebout's inequality (a refinement of the Cauchy-Schwarz inequality) \cite{CALLEBAUT1965491} shows that the subnormalisation $α_p$ improves as $p\to 1/2$:
\begin{equation}
    \frac{\sqrt{S_cS_r}}{D}\sum_d |A_d| = α_{1/2} \le α_p \le α_q \ \text{for}\ 1/2 \le p \le q.
    \label{eq:subnormalisation prep inequality}
\end{equation}
There is a single optimal choice $p=1/2$ of the parameter for all matrices, contrary to preamplification. In fact, for $p=1/2$, one can choose $\text{UNPREP}=\text{PREP}^\dag$ up to $\text{sgn}(A_d)$, and for certain circuit implementations of $\text{PREP}$ one can use measurement based uncomputation to significantly reduce the number of Toffoli gates in $\text{UNPREP}$ \cite{berry_qubitization_2019}. Effectively, we take this into account by recording reduced data loading $D$ for the optimal $p=1/2$ in table~\ref{table:summary}.

Compared to our base scheme, the subnormalisation is always reduced:
\begin{equation}
α_p = \sqrt{S_cS_r}\sqrt{\frac{\sum_d |A_d|^{2p}}{D}}\sqrt{\frac{\sum_d|A_d|^{2-2p}}{D}} \le \sqrt{S_cS_r} ||A||_\text{max}^p ||A||_\text{max}^{1-p} = \sqrt{S_cS_r}||A||_\text{max}.
\end{equation}
The reduction is stronger when the matrix values have large varying magnitude than if they are all of similiar size.
Data loading cost is only increased by a factor of 2 compared to the base scheme; except for the optimal $p=1/2$, when it is the same.
Whenever possible (i.e.~$D\le S_c,S_r$ and the row and column oracles commute with the data loading), one should therefore choose PREP/UNPREP with $p=1/2$ instead of the base scheme.
%
It depends on the specific matrix including its values whether the preamplified or PREP/UNPREP scheme have a better figure of merit \eqref{eq:figure of merit}.

\subsection{Hermitian block encoding for symmetric matrices}
\label{sec:hermitian}
When the matrix $A$ is symmetric, it can be desirable to construct a Hermitian block encoding. That simplifies quantum walks \cite{szegedy_quantum_2004,berry_hamiltonian_2015} and phase estimation via qubitisation \cite{low_hamiltonian_qubitisation_2019}. 
A priori, the block encodings constructed above may not be Hermitian, even if the matrix is symmetric.

In this section, we show how all our block encodings become Hermitian with only slight modifications when the oracles are constructed in a particular way. The starting point is, like in all our constructions, a labelling $(d,m)$ of the matrix elements, and a column oracle $O_c: (d,m) \mapsto (j,s_c)$. Because the matrix is symmetric, elements related by transposition ($i\leftrightarrow j$) have the same $d$. Thus, there is an oracle $O_t$ that maps $(d,m) \mapsto (d,m')$, the element related to $(d,m)$ by transposition. On the diagonal of the matrix, $m'=m$. When the arithmetic expression is cast into a quantum circuit, it becomes unitary and Hermitian, because transposition is an involution $O_t^2=\mathbb{I}$. As usual, the action of $O_t$ on out-of-range $(d,m)$ does not matter (apart from ensuring its unitarity and Hermiticity) because of $O_\text{rg}$ flipping the delete qubit.
A row oracle can then be constructed from the column oracle as
\begin{equation}
    O_r = O_c O_t,
    \label{eq:O_r from O_t}
\end{equation}
keeping in mind that $S_c=S_r$. We will show examples of constructions of Hermitian block encodings in section~\ref{sec:examples}. A further $Z$ gate must be added to the data qubit to make the base block encoding scheme Hermitian:
\begin{equation}
\begin{tikzpicture}
\begin{yquant}
    qubit {data} f0;
    qubit del;
    qubit {s} f2;
    qubit {$\ket{j}$} block;
    
    ["north east:$S$" {font=\protect\footnotesize, inner sep=0pt}]
    slash f2;
    ["north east:$N$" {font=\protect\footnotesize, inner sep=0pt}]
    slash block;
    hspace {5pt} -;
    box {$H_{S_c}$} f2;
    
    align -;
    text {$s_c$} f2;
    text {$j$} block;
    box {$O_c^\dag$} (f2, block);
    setstyle {thick, riverlane_pink} f2, block;
    text {$d$} f2;
    text {$m$} block;
    hspace {-3pt} -;
    ["north east:$D$" {font=\protect\footnotesize, inner sep=0pt}]
    slash f2;
    ["north east:$M$" {font=\protect\footnotesize, inner sep=0pt}]
    slash block;
    hspace {5pt} -;

    [plusctrl, shape=yquant-circle]
    box {$O_\text{rg}$} (block, f2) | del;

    z f0;
    [multictrl]
    box {$R(A_d/||A||_\text{max})$} f0 ~ f2;

    box {$O_t$} (block, f2);
    text {$d$} f2;
    text {$m$} block;
    box {$O_c$} (f2, block);
    setstyle {black} f2, block;
    text {$s_r$} f2;
    text {$i$} block;
    
    box {$H_{S_c}^\dag$} f2;
    output {$\ket{i}$} block;
\end{yquant}
\end{tikzpicture}.
\label{eq:hermitian block encoding circuit}
\end{equation}
The circuit is Hermitian, because the left and right parts are Hermitian conjugates, and the middle part consists of three commuting Hermitian operators (the $O_\text{rg}$-controlled NOT, $O_t$, and the $X$-axis rotations \eqref{eq:formula multiplexed rotations} preceded by $Z$). The $Z$ does not affect the encoded matrix because it has no effect when the flag qubit is initialised as $\ket{0}$. Yet, it is needed to make the full block encoding unitary Hermitian.
This Hermitian counterpart of the base block encoding scheme has the same data loading cost, subnormalisation, and flag qubit number. 

The preamplified scheme can also be made Hermitian by using the $O_r$ oracle \eqref{eq:O_r from O_t} constructed above. Using $p=1/2$, the circuit \eqref{eq:circuit preamplification split} can be adapted to:
\begin{equation}
\begin{tikzpicture}[scale=0.89]
\begin{yquant}
    qubit {data0} f0a;
    qubit {data1}f0b;
    qubit {del} f1;
    qubit {s} f2;
    qubit {$\ket{j}$} block;
    [register/minimum height=15pt]
    
    ["north east:$S$" {font=\protect\footnotesize, inner sep=0pt}]
    slash f2;
    ["north east:$N$" {font=\protect\footnotesize, inner sep=0pt}]
    slash block;
    hspace {8pt} -;
    
    [name=leftD]
    box {$H_{S_c}$} f2;
    
    align -;
    text {$s_c$} f2;
    text {$j$} block;
    [name=Oc]
    box {$O_c^\dag$} (f2, block);
    setstyle {thick, riverlane_pink} f2, block;
    text {$d$} f2;
    text {$m$} block;
    hspace {-3pt} -;
    ["north east:$D$" {font=\protect\footnotesize, inner sep=0pt}]
    slash f2;
    ["north east:$M$" {font=\protect\footnotesize, inner sep=0pt}]
    slash block;
    hspace {-25pt} -;
    [multictrl, name=leftR]
    box {$R\left(\frac{\sqrt{|A_d|}}{\max_{d'} \sqrt{|A_{d'}|}}\right)$} f0b ~ f2;
    [name=leftZ]
    z f0b;

    hspace {5pt} -;
    
    [plusctrl, shape=yquant-circle]
    box {$O_\text{rg}$} (f2, block) | f1;
    swap (f0a, f0b);
    box {$O_t$} (f2, block);
    
    box {$\sgn A_d$} f2;

    hspace {5pt} -;

    [multictrl, name=rightR]
    box {$R\left(\frac{\sqrt{|A_d|}}{\max_{d'} \sqrt{|A_{d'}|}}\right)$} f0b ~ f2;
    [name=rightZ]
    z f0b;
    hspace {-35pt} -;
    align -;
    text {$d$} f2;
    text {$m$} block;
    [name=Or]
    box {$O_c$} (f2, block);
    setstyle {black} f2, block;
    text {$s_r$} f2;
    text {$i$} block;
    [name=rightD]
    box {$H_{S_r}^\dag$} f2;
    hspace {5pt} -;
    output {$\ket{i}$} block;
    
    \node[draw, dashed, fit=(leftD) (leftR) (Oc) (leftZ), "$U_c^\dag$" below] {};
    \node[draw, dashed, fit=(rightD) (rightR) (Or) (rightZ), "$U_c$" below] {};
\end{yquant}
\end{tikzpicture}
\label{eq:circuit Hermitian preamplification split}
\end{equation}
By removing the sign of $A_d$ from the first multiplexed rotations in \eqref{eq:circuit preamplification split}, the two subcircuits to be amplified can be made Hermitian conjugates. The same will hold for their amplification circuits. The correct sign is obtained with the gate denoted by $\sgn A_d$. It can be seen as a new data loading oracle loading and applying the sign. It can be implemented with multicontrolled $Z$ gates flipping the signs for those $\ket{d}$ with $\sgn A_d=-1$. All circuit elements in the middle are Hermitian and commute. Compared to the preamplification scheme, this Hermitian counterpart has the same number of flag qubits, the same subnormalisation, but data loading increased by $D$, due to the separate sign oracle.

The PREP/UNPREP scheme is already Hermitian as-is for $p=1/2$, provided $O_r=O_cO_t$ is used for the row oracle and matching $\text{PREP}$ and $\text{UNPREP}$ operators are used: While equations \eqref{eq:prep} and \eqref{eq:unprep} only specify the operators' actions on $\ket{0}$, the identity
\begin{equation}
    \text{PREP} = O_\text{sgn}\text{UNPREP}^\dagger 
\end{equation}
must hold as an operator identity to make the full circuit Hermitian. The oracle $O_\text{sgn}$ flips the signs to match $\sgn(A_d)$. Because $O_\text{sgn}$ is Hermitian, the full PREP/UNPREP block encoding \eqref{eq:circuit prep block encoding} is Hermitian. In practice, $O_\text{sgn}$ is integrated into PREP, and the Hermitian counterpart of the PREP/UNPREP scheme has the same costs.

A simple way to make any block encoding $U$ Hermitian is the following circuit \cite{low_hamiltonian_qubitisation_2019}:
\begin{equation}
\begin{tikzpicture}
\begin{yquant}
qubit {new flag qubit} s;
qubit {flag qubits} flags;
qubit {$\ket{j}$} block;
slash flags;
slash block;
align -;
h s;
box {$U$} (flags, block) | s;
x s;
box {$U^\dag$} (flags, block) | s;
h s;
output {$\ket{i}$} block;
\end{yquant}
\end{tikzpicture}.
\end{equation}
It increases the number of flag qubits by 1 and increases the gate complexity / data loading cost by a factor of two. Additional cost is incurred for controlling $U$ and $U^\dag$. 

A Hermitian block encoding scheme in terms of certain black-box oracles is suggested in  \cite{camps_explicit_2022}. It cannot deal with negative matrix values and has $\log_2 N + 2$ flag qubits. It also duplicates data loading cost, in contrast to our Hermitian counterpart of the base block encoding.

\section{Example block encodings}
\label{sec:examples}

\subsection{Checkerboard matrix}
We will first consider a matrix with a checkerboard pattern as a nice example demonstrating how our block encoding scheme provides for repeated values, and application of PREP/UNPREP.
The $N\times N$ checkerboard matrix (take $N$ a power of 2) only has $D=2$ data values, the sparsities are $S_c=S_r=S=N$ (it is dense), and $M=N^2/2$ multiplicity of each of the two values.
It clearly has an arithmetic structure, which can be translated into row and column oracles.
The starting point is a labelling of the matrix elements by $(d,m)$, $0\le d < D$, $0\le m < M$. We choose a labelling where $m$ increases reading the matrix row by row from left to right. Specifically, for $N=4$, we have the labelling:
\begin{equation}
\begin{pmatrix}
(d=0, m=0) & (d=1, m=0) & (d=0, m=1) & (d=1, m=1) \\
(d=1, m=2) & (d=0, m=2) & (d=1, m=3) & (d=0, m=3) \\
(d=0, m=4) & (d=1, m=4) & (d=0, m=5) & (d=1, m=5) \\
(d=1, m=6) & (d=0, m=6) & (d=1, m=7) & (d=0, m=7)
\end{pmatrix}
\end{equation}

For simplicity of presentation we split $m = N m^\text{hi} + (N/2) m^\text{mid} + m^\text{lo}$ into its high $\log(N/2)$ bits, mid bit, and low $\log(N/2)$ bits.
The row and column indices $i$ and $j$ can be computed from $(d,m)$ as follows:
\begin{align}
i(d,m) &= \left\lfloor \frac{m}{N/2} \right\rfloor = 2m^\text{hi} + m^\text{mid}\\
j(d,m) &= 2 (m\ \text{mod}\ (N/2)) + (d+i\ \text{mod}\ 2) = 2m^\text{lo} + (d+m^\text{mid}\ \text{mod}\ 2).
\end{align}
The specific values of $s_c$ and $s_r$ are not important for the block encoding.
Oracles for these expressions are very simple, $O_r$ can even be implemented as the identity. The full block encoding circuit is:
\begin{equation}
\begin{tikzpicture}
\begin{yquant}
qubit {} f0;
qubit {} g;
qubit {} hlo;
qubit {} block;
qubit {} block0;
init {$\ket{j}$} (block, block0);

["north east:$N/2$" {font=\protect\footnotesize, inner sep=0pt}]
slash hlo;
["north east:$N/2$" {font=\protect\footnotesize, inner sep=0pt}]
slash block;
hspace {10pt} -;

box {$H_{N}$} (hlo,g);
hspace {1pt} -;
init {$s_c$} (g,hlo);

[this subcircuit box style={dashed, "$O_c^\dag$" below}]
subcircuit {
    qubit {} g;
    qubit {} hlo;
    qubit {} block;
    qubit {} block0;

    swap (hlo, block);
    cnot block0 | g;

} (g,hlo,block,block0);

    inspect {$d$} g;
    inspect {$m^\text{lo}$} hlo;
    inspect {$m^\text{mid}$} block0;
    inspect {$m^\text{hi}$} block;
[multictrl]
box {$R(A_d/||A||_\text{max})$} f0 ~ g;

[this subcircuit box style={dashed, "$O_r$" below}]
subcircuit {
    qubit {} g;
    qubit {} hlo;
    qubit {} block;
    qubit {} block0;
    hspace {10pt} -;
} (g,hlo,block,block0);

inspect {$s_r$} (g, hlo);
hspace {1pt} -;
box {$H_{N}^\dag$} (hlo,g);
output {$\ket{i}$} (block, block0);
\end{yquant}
\end{tikzpicture}
\label{eq:checkerboard base circuit}
\end{equation}

The block encoding has data loading cost $D=2$. Note that the oracles are very cheap, in fact, zero $T$ gate count. It turns out that this block encoding is already Hermitian.
The subnormalisation is
\begin{equation}
    \sqrt{S_rS_c}||A||_\text{max} = N\max\left(|A_0|,|A_1|\right)
    \label{eq:subnormalisation checkerboard base}
\end{equation}

The subnormalisation can be improved by going to PREP/UNPREP, since $D\le S$ and the oracles commute with the multiplexed rotations. The corresponding block encoding circuit is:
\begin{equation}
\begin{tikzpicture}
\begin{yquant}
qubit {} d;
qubit {} lo;
qubit {} hi;
qubit {} mid;
init {$\ket{j}$} (hi, mid);

["north east:$N/2$" {font=\protect\footnotesize, inner sep=0pt}]
slash lo;
["north east:$N/2$" {font=\protect\footnotesize, inner sep=0pt}]
slash hi;
hspace {10pt} -;

box {PREP} d;
box {$H_{N/2}$} lo;
swap (lo,hi);
cnot mid | d;
box {UNPREP} d;
box {$H_{N/2}^\dag$} lo;
output {$\ket{i}$} (hi,mid);
\end{yquant}    
\end{tikzpicture}
\label{eq:checkerboard prep circuit}
\end{equation}
with
\begin{equation}
    \text{PREP}\ket{0} = \frac{1}{\sqrt{|A_0|+|A_1|}}\left( \sqrt{A_0}\ket{0} + \sqrt{A_1}\ket{1}\right).
\end{equation}
The subnormalisation follows as
\begin{equation}
    \frac{\sqrt{S_cS_r}}{D} \sum_d |A_d| = N\frac{|A_0|+|A_1|}{2}.
\end{equation}
When the matrix elements have varying magnitude, this presents an improvement over the subnormalisation \eqref{eq:subnormalisation checkerboard base} of the base scheme.

The checkerboard matrix serves as a simple pedagogical example. Because of its low rank and known eigenvalues, it is unlikely to be used in an actual quantum algorithm. Alternatively to our schemes, a block encoding could also be constructed by appreciating the factorisation
\begin{equation}
A = \left(\ket{+}\!\!\bra{+}\right)^{\otimes \log_2 N-1} \otimes \left(A_0\mathbb{I} +A_1X\right)
\end{equation}
of the matrix, leading to a circuit with the same ancilla, $T$ count and subnormalisation as our schemes.

Our scheme can adapt to changes in the structure of the matrix. For example, to demonstrate the out-of-range $O_\text{rg}$,  consider a checkerboard matrix with top left and bottom right entries replaced by zero. The row and column oracles are as above; the additional $O_\text{rg}$ oracle deletes out-of-range labels $(d,m)$. These are $(d=0,m=0)$ and $(d=0, m=N^2/2-1)$ for top-left and bottom-right entries. A circuit implementation of the oracle is
\begin{equation}
\begin{tikzpicture}
\begin{yquant}
qubit del;
qubit {$d$} d;
qubit {$m^\text{lo}$} lo;
qubit {$m^\text{hi}$} hi;
qubit {$m^\text{mid}$} mid;

["north east:$N/2$" {font=\protect\footnotesize, inner sep=0pt}]
slash lo;
["north east:$N/2$" {font=\protect\footnotesize, inner sep=0pt}]
slash hi;
hspace {8pt} -;

[plusctrl, shape=yquant-circle]
box {$O_\text{rg}$} (d, lo,hi,mid) | del;
text {$=$} (-);
cnot del | hi, lo, mid ~ d;
cnot del ~ d, hi, lo, mid;
\end{yquant}
\end{tikzpicture}
\end{equation}
and can be inserted in both the base circuit \eqref{eq:checkerboard base circuit} and the PREP/UNPREP circuit \eqref{eq:checkerboard prep circuit}.

\subsection{Toeplitz matrix}
\label{sec:toeplitz}
Consider a Toeplitz matrix with $D$ diagonals, i.e.~$D$ values, offset from the main diagonal by $k$:
\begin{equation}
\begin{pmatrix}
A_{k} & A_{k-1} & \cdots & A_0 & & & & \\
A_{k+1} & A_{k} & A_{k-1} & \cdots & A_0 & & & \\
\vdots      &  A_{k+1} & A_{k} & A_{k-1} & \cdots & A_{0} & & \\
A_{D-1} & \vdots & A_{k+1} & A_{k} & A_{k-1} & \cdots & A_{0} & \\
& A_{D-1} & \vdots & A_{k+1} & A_{k} & A_{k-1} & \cdots & A_{0} \\
& & A_{D-1} & \vdots & A_{k+1} & A_{k} & A_{k-1} & \cdots \\
& & & A_{D-1} & \vdots & A_{k+1} & A_k & A_{k-1} \\
& & & & A_{D-1} & \vdots & A_{k+1} & A_k \\
\end{pmatrix}
\label{eq:toeplitz matrix}
\end{equation}
For an $N\times N$ matrix with $N\ge D$ it follows $S_c=S_r=S=D$, $M=N$. This already fulfills $NS_c=NS_r=MD$ such that no padding is necessary, and we assume $N$ and $D$ to be powers of 2 for simplicity.

Different choices can be made for the labelling in terms of $(d,m)$ (distinct values and their repetitions). For the distinct values, we choose $d$ as above. For $m$, we will simply choose the column index. Note that for example, ($d=0$, $m=0$) is out-of-range: In general, there is no $A_0$ in column 0.
Arithmetically, the mapping to row and column indices is then:
\begin{equation}
    i(d,m) = d - k + m,\quad j(d,m) = m
    \label{eq:ij toeplitz}
\end{equation}
and out-of-range $(d,m)$ pairs are those where an overflow or underflow in the calculation of $i$ occurs, that is when
\begin{equation}
    d-k+m < 0 \ \text{or}\ d-k+m \ge N.
\end{equation}

The oracles can be constructed with in-place additions, such that the block-encoding circuit is as follows.  The last qubit is an ancilla qubit serving as the overflow (as the next higher bit) for the modular additions. It is not a flag qubit because it is uncomputed back to $\ket{0}$. (Note the additions can be performed without further ancillas if only few qubits are available \cite{draper_addition_2000,cuccaro_new_2004,ruiz-perez_quantum_2017}.)
\begin{equation}
\begin{tikzpicture}
\begin{yquant}
qubit data;
qubit del;
qubit {s} g;
qubit {$\ket{j}$} block;
qubit overflow;
["north east:$D$" {font=\protect\footnotesize, inner sep=0pt}]
slash g;
["north east:$N$" {font=\protect\footnotesize, inner sep=0pt}]
slash block;
hspace {5pt} -;

box {$H_D$} g;

[this subcircuit box style={dashed, "$O_c^\dag$" below}]
subcircuit {
    qubit {} q[2];
    hspace {10pt} -;
} (g,block);
hspace {3pt} -;
text {$d$} g;
text {$m$} block;
hspace {6pt} -;
    [multictrl,name=ovstart]
    box {$+d$} (block,overflow) ~ g;
    box {$-k$} (block,overflow);
    cnot del | overflow;
    box {$+k$} (block,overflow);
    [multictrl,name=ovend]
    box {$-d$} (block,overflow) ~ g;

[multictrl]
box {$R(A_d/||A||_\text{max})$} data ~ g;

[this subcircuit box style={dashed, "$O_r$" below}]
subcircuit {
    qubit {} g;
    qubit {} block;
    [multictrl]
    box {$+d$} block ~ g;
    box {$-k$} block;
} (g,block);

box {$H_D^\dag$} g;
output {$\ket{i}$} block;
\node[draw, dashed, fit=(ovstart) (ovstart-n0) (ovend), "$O_\text{rg}$" below] {};
\end{yquant}
\end{tikzpicture}
\end{equation}
If we were encoding a banded circulant matrix, the equations \eqref{eq:ij toeplitz} would have be taken mod $N$. There would be no out-of-range pairs $(d,m)$ and no $O_\text{rg}$ oracle, del qubit, or overflow ancilla qubit. The above circuit would then be a block encoding with $\log_2 D +1 =\log_2 S+1$ flag qubits, subnormalisation $S||A||_\text{max}$, and $D$ data loading. That is the same as the block encoding discussed in \cite{camps_explicit_2022}.

Yet for the Toeplitz matrix, the circuit can be simplified to be more compact. The overflow qubit can be used directly as the del flag qubit; then only the modular part of the addition/subtraction must be uncomputed:
\begin{equation}
\begin{tikzpicture}
\begin{yquant}
qubit data;
qubit {s} g;
qubit {$\ket{j}$} block;
qubit {del} overflow;
["north east:$D$" {font=\protect\footnotesize, inner sep=0pt}]
slash g;
["north east:$N$" {font=\protect\footnotesize, inner sep=0pt}]
slash block;
hspace {5pt} -;

box {$H_D$} g;

[this subcircuit box style={dashed, "$O_c^\dag$" below}]
subcircuit {
    qubit {} q[2];
    hspace {10pt} -;
} (g,block);
hspace {6pt} -;
text {$d$} g;
text {$m$} block;
hspace {10pt} -;
    [multictrl,name=ovstart]
    box {$+d$} (block,overflow) ~ g;
    box {$-k$} (block,overflow);

    box {$+k$} (block);
    [multictrl,name=ovend]
    box {$-d$} (block) ~ g;

[multictrl]
box {$R(A_d/||A||_\text{max})$} data ~ g;

[this subcircuit box style={dashed, "$O_r$" below}]
subcircuit {
    qubit {} g;
    qubit {} block;
    [multictrl]
    box {$+d$} block ~ g;
    box {$-k$} block;
} (g,block);

box {$H_D^\dag$} g;
output {$\ket{i}$} block;
\node[draw, dashed, fit=(ovstart) (ovstart-n0) (ovend), "$O_\text{rg}$" below] {};
\end{yquant}
\end{tikzpicture}
\end{equation}
Then, the $O_\text{rg}$ and $O_r$ oracles can be merged, resulting in the simpler circuit
\begin{equation}
\begin{tikzpicture}
\begin{yquant}
    qubit {data} f0;
    qubit {s} g;
    qubit {$\ket{j}$} block;
    qubit del;
    
    ["north east:$D$" {font=\protect\footnotesize, inner sep=0pt}]
    slash g;
    ["north east:$N$" {font=\protect\footnotesize, inner sep=0pt}]
    slash block;
    hspace {5pt} -;

    box {$H_D$} g;
    
    [multictrl]
    box {$R(A_d/||A||_\text{max})$} f0 ~ g;

    [multictrl]
    box {$+d$} (block,del) ~ g;
    box {$-k$} (block,del);

    box {$H_D^\dag$} g;
    output {$\ket{i}$} block;
\end{yquant}
\end{tikzpicture}.
\end{equation}
The circuit can also be interpreted as block-encoding a circulant matrix, and the del qubit selecting the top left block: Every Toeplitz matrix can be embedded in a larger circulant matrix. Such an observation was already used for HHL \cite{mahasinghe_efficient_2016}. Further, the Fourier transformation of a circulant matrix is diagonal. Conceivably, one could construct a block encoding of the diagonal Fourier transformation. That approach is equivalent to compiling the additions in the above circuit with QFT based adders \cite{draper_addition_2000}.

The arithmetics in the oracles only have $\propto\log N$ Toffoli gates. The flag qubit number is $2+\log_2 D$ and we have $D$ data loading.
The Gily\'{e}n et.~al.~method would have more flag qubits, $3 + \log_2 N$.

Note that this matrix has $D\le S$, and the row and column oracles commute with the data loading oracle. Hence, application of PREP/UNPREP can reduce the subnormalisation, resulting in the circuit
\begin{equation}
\begin{tikzpicture}
\begin{yquant}
qubit {} d;
qubit {$\ket{j}$} block;
qubit del;

["north east:$D$" {font=\protect\footnotesize, inner sep=0pt}]
slash d;
["north east:$N$" {font=\protect\footnotesize, inner sep=0pt}]
slash block;
hspace {5pt} -;

box {PREP} d;

[multictrl]
box {$+d$} (block,del) ~ d;
box {$-k$} (block,del);

box {UNPREP} d;
output {$\ket{i}$} block;
\end{yquant}
\end{tikzpicture}.
\label{eq:toeplitz prep circuit}
\end{equation}

\subsection{Tridiagonal matrix}

We consider a matrix that is tridiagonal and symmetric, where entries along a diagonal are all different. For an $N\times N$ matrix (we consider $N$ a power of 2), we have $D=2N-1$, $M=2$, $S_c=S_r=S=3$.
A possible labelling in terms of $(d,m)$ is
\begin{equation}
\begin{pmatrix}
(d=0, m=0) & (d=1, m=0) & & & & \\
(d=1, m=1) & (d=2, m=0) & (d=3, m=0) & & & \\
& (d=3, m=1) & (d=4,m=0) & \ddots & & \\
& & \ddots & \ddots &  (d=2N-3, m=0) \\
& & & (d=2N-3, m=1) & (d=2N-2, m=0) \\
\end{pmatrix}.
\end{equation}
Since $SN=3N\neq DM = 4N-4$, we must pad $S$ and/or $D$. The equality can be fullfilled by padding to $D=2N$ and $S=4$.

Arithmetic expressions for row and column indices can be written as
\begin{align}
i(d,m) &= \left\lfloor \frac{d}{2} \right\rfloor + m = d^\text{hi} + m\\
j(d,m) &= \left\lfloor \frac{d}{2}\right\rfloor + \{d\ \text{mod}\ 2\ \text{if}\ m=0\} = d^\text{hi} + \{d^\text{lo}\ \text{if}\ m=0\}
\end{align}
when splitting $d=2d^\text{hi} + d^\text{lo}$ into its high $\log_2 N$ bits and lowest bit.
Pairs $(d,m)$ out-of-range are those with
\begin{equation}
    (d^\text{lo} =0\ \text{and}\ m=1)\ \text{or}\ d=2N-1.
\end{equation}
Translating these expressions to quantum circuits results in the block encoding circuit
\begin{equation}
\begin{tikzpicture}
\begin{yquant}
qubit data;
qubit del;
qubit {s1} h;
qubit {s0} g0;
qubit {$\ket{j}$} grest;
["north east:$N$" {font=\protect\footnotesize, inner sep=0pt}]
slash grest;
hspace {5pt} -;
box {$H_3$} (h,g0);
hspace {3pt} -;

[this subcircuit box style={dashed, "$O_c^\dag$" below}]
subcircuit {
    qubit {} h;
    qubit {} g0;
    qubit {} grest;
    box {$-1$} grest | g0 ~ h;
} (h,g0,grest);

text {$m$} h;
text {$d^\text{lo}$} g0;
text {$d^\text{hi}$} grest;
align -;
[name=ov1]
cnot del | h ~ g0;
[name=ov2]
cnot del | g0, grest;

[multictrl]
box {$R(A_d/||A||_\text{max})$} data ~ g0, grest;

[this subcircuit box style={dashed, "$O_r$" below}]
subcircuit {
    qubit {} h;
    qubit {} g0;
    qubit {} grest;
    box {$+1$} grest | h;
} (h,g0,grest);
hspace {3pt} -;
box {$H_3^\dag$} (h,g0);
output {$\ket{i}$} grest;

\node[draw, dashed, fit=(ov1) (ov2) (ov2-p1), "$O_\text{rg}$" below] {};
\end{yquant}
\end{tikzpicture}.
\end{equation}
The equal superposition prepared by $H_3$ must be 
\begin{equation}H_3\ket{0}^{\text{s}_1}\ket{0}^{\text{s}_0} = (\ket{0}^{\text{s}_1}\ket{0}^{\text{s}_0}+\ket{0}^{\text{s}_1}\ket{1}^{\text{s}_0}+\ket{1}^{\text{s}_1}\ket{1}^{\text{s}_0})/\sqrt{3},
\label{eq:H3_tridiagonal}\end{equation}
since those are the values of $s_c$ and $s_r$ that the oracles map valid $(d,m)$ pairs to. Note that in this example, $S>D$ and the PREP/UNPREP method cannot be applied. Indeed, the data loading does not commute with the oracles.

Note the Tofolli count of the arithmetic oracles is logarithmic in $N$, and the addition and subtraction can be implemented without further ancilla qubits. The Toffoli cost is dominated by data loading of $D=2N$ items. There are only $4$ flag qubits. In the Gily\'{e}n et.~al.~scheme, there would be $3+\log_2 N$ flag qubits.

The circuit can be simplified further by appreciating that the first CNOT in $O_\text{rg}$ will never be triggered (it commutes with the column and row oracles and is annihilated by the $H_3$ of eq.~\eqref{eq:H3_tridiagonal}) and can be removed. Further, instead of the second multicontrolled NOT in $O_\text{rg}$, one could explicitly load a zero value for $d=2N-1$ with the multiplexed rotations. Then the block encoding does not need a delete flag qubit or $O_\text{rg}$ oracle:
\begin{equation}
\begin{tikzpicture}
\begin{yquant}
qubit data;
qubit {s1} h;
qubit {s0} g0;
qubit {$\ket{j}$} grest;
["north east:$N$" {font=\protect\footnotesize, inner sep=0pt}]
slash grest;
hspace {5pt} -;
box {$H_3$} (h,g0);
hspace {3pt} -;

[this subcircuit box style={dashed, "$O_c^\dag$" below}]
subcircuit {
    qubit {} h;
    qubit {} g0;
    qubit {} grest;
    box {$-1$} grest | g0 ~ h;
} (h,g0,grest);

text {$m$} h;
text {$d^\text{lo}$} g0;
text {$d^\text{hi}$} grest;
align -;

[multictrl]
box {$R(A_d/||A||_\text{max})$} data ~ g0, grest;

[this subcircuit box style={dashed, "$O_r$" below}]
subcircuit {
    qubit {} h;
    qubit {} g0;
    qubit {} grest;
    box {$+1$} grest | h;
} (h,g0,grest);
hspace {3pt} -;
box {$H_3^\dag$} (h,g0);
output {$\ket{i}$} grest;
\end{yquant}
\end{tikzpicture},
\end{equation}
with $A_{2N-1} = 0$.

The above block encoding circuit is not Hermitian. We will demonstrate how to use our scheme to construct a Hermitian block encoding.
The $O_t$ oracle has the following action:
\begin{equation}
O_t(d,m) = \begin{cases}(d,m) &\text{for $d$ even}\\
(d, \text{NOT}\ m) &\text{for $d$ odd}\end{cases}
\end{equation}
and can easily be implemented in a quantum circuit:
\begin{equation}
\begin{tikzpicture}
\begin{yquant}
qubit {$m$} m;
qubit {$d^\text{lo}$} dlo;
qubit {$d^\text{hi}$} dhi;
["north east:$N$" {font=\protect\footnotesize, inner sep=0pt}]
slash dhi;
hspace {6pt} -;
box {$O_t$} (-);
text {$=$} (-);
cnot m | dlo;
\end{yquant}
\end{tikzpicture}
\end{equation}
The oracle column oracle $O_c$ is the same as above, and the row oracle is now constructed as $O_r =O_cO_t$. Putting this into the Hermitian version of the base block encoding circuit \eqref{eq:hermitian block encoding circuit} gives
\begin{equation}
\begin{tikzpicture}
\begin{yquant}
qubit data;
qubit s1;
qubit s0;
qubit {$\ket{j}$} block;
["north east:$N$" {font=\protect\footnotesize, inner sep=0pt}]
slash block;
align -;
box {$H_3$} (s0,s1);
align -;
box {$-1$} block | s0 ~ s1;
z data;
[multictrl]
box {$R(A_d/||A||_\text{max})$} data ~ s0, block;
cnot s1 | s0;
box {$+1$} block | s0 ~ s1;
box {$H_3^\dag$} (s0,s1);
output {$\ket{i}$} block;
\end{yquant}
\end{tikzpicture}.
\end{equation}

\subsection{Extended binary tree matrix}
The symmetric adjacency matrix of a balanced binary tree was considered in \cite{camps_explicit_2022}. While the authors considered a non-Hermitian block encoding for $N=8$, we will demonstrate how our scheme can be used to generate a Hermitian block encoding for $N$ an arbitrary power of 2. For $N=8$, the binary tree and its adjacency matrix are
\begin{equation}
\begin{tikzpicture}[level distance=0.8cm,
    level 2/.style={sibling distance=3cm},
    level 3/.style={sibling distance=1.5cm},
    baseline=(current bounding box.center)]
\node {0}
    child { node {1}
            child { node {2}
                    child { node {4} }
                    child { node {5} }
            }
            child { node {3}
                    child { node {6} }
                    child { node {7} }
            }
    };
\end{tikzpicture}\qquad
\begin{pmatrix}
A_0 & A_2 & & & & & & \\
A_2 & A_1 & A_2 & A_2 & & & & \\
 & A_2 & A_1 & & A_2 & A_2 & & \\
 & A_2 & & A_1 & & & A_2 & A_2 \\
 &  & A_2 & & A_0 & & & \\
 &  & A_2 &  & & A_0 & & \\
 &  &  & A_2 & & & A_0 & \\
 &  &  & A_2 & & &  & A_0 
\end{pmatrix}.
\end{equation}
The root and leaf nodes have weight $A_0$, all other nodes weight $A_1$, and the edges weight $A_2$. The labelling of $d$ is already exemplified in above matrix.
One can see that $D=3, M=2N-2, S=S_r=S_c=4$. Hence, $M$ is padded to $2N$, and $S$ is padded to $6$. Moreover, in this example $D$ and the padded $S$ are not powers of two, such that they are embedded in 2 and 3 qubits, respectively.

The labelling of $m$ can be chosen as
\begin{equation}
\begin{pmatrix}
0 & 1 & & & & & & \\
9 & 1 & 2 & 3 & & & & \\
    & 10 & 2 & & 4 & 5 & & \\
    & 11 & & 3 & & & 6 & 7 \\
    &  & 12 & & 4 & & & \\
    &  & 13 &  & & 5 & & \\
    &  &  & 14 & & & 6 & \\
    &  &  & 15 & & &  & 7
\end{pmatrix},
\end{equation}
which generalises to the following relations:
\begin{align}
i(d,m) &= \begin{cases}m & \text{for}\ d=0,1\\
    \left\lfloor \frac{m}{2}\right\rfloor & \text{for}\ d=2\ \text{and}\ m<N \\
    m-N & \text{for}\ d=2\ \text{and}\ m>N\end{cases} \label{eq:i tree}\\
m'(d,m) &= \begin{cases}m & \text{for}\ d=0,1\\
    m+N& \text{for}\ d=2\ \text{and}\ m<N\\
    m-N& \text{for}\ d=2\ \text{and}\ m>N\\\end{cases}
\end{align}
with out-of-range $(d,m)$ those with $(d=0,m\ge N)$ or $(d=0, 1\le m < N/2)$ or $(d=1,m=0)$ or $(d=1,m\ge N/2)$ or $(d=2, m=0,N)$ or $d=3$.

The oracle $O_t$ can be easily implemented by flipping one qubit because we are assuming that $N$ is a power of 2.
\begin{equation}
\begin{tikzpicture}
\begin{yquant}
qubit {$d^\text{hi}$} dhi;
qubit {$d^\text{lo}$} dlo;
qubit {$m^\text{hi}$} mhi;
qubit {$m^\text{mid}$} mmid;
qubit {$m^\text{lo}$} mlo;
["north:$N/2$" {font=\protect\footnotesize, inner sep=0pt}]
slash mlo;
align -;
box {$O_t$} (-);
hspace {3pt} -;
text {\ =\ } (-);

init {$d^\text{hi}$} dhi;
init {$d^\text{lo}$} dlo;
init {$m^\text{hi}$} mhi;
init {$m^\text{mid}$} mmid;
init {$m^\text{lo}$} mlo;
cnot mhi | dhi;
output {$d^\text{hi}$} dhi;
output {$d^\text{lo}$} dlo;
output {${m'}^\text{hi}$} mhi;
output {${m'}^\text{mid}$} mmid;
output {${m'}^\text{lo}$} mlo;
\end{yquant}
\end{tikzpicture}.
\end{equation}
When implementing the oracle $O_r$, we must make sure to only get $S_r=3$ values of $s_r$ for valid $(d,m)$.
\begin{equation}
\begin{tikzpicture}
\begin{yquant}
qubit {$d^\text{hi}$} dhi;
qubit {$d^\text{lo}$} dlo;
qubit {$m^\text{hi}$} mhi;
qubit {$m^\text{mid}$} mmid;
qubit {$m^\text{lo}$} mlo;
["north:$N/2$" {font=\protect\footnotesize, inner sep=0pt}]
slash mlo;
align -;
box {$O_r$} (-);
hspace {3pt} -;
text {\ =\ } (-);
init {$d^\text{hi}$} dhi;
init {$d^\text{lo}$} dlo;
init {$m^\text{hi}$} mhi;
init {$m^\text{mid}$} mmid;
init {$m^\text{lo}$} mlo;

x dlo ~ dhi;
cnot dlo ~ dhi, mmid, mlo;
cnot dlo | mmid ~ dhi;
text {$\ket{0}$} dlo;
cnot dlo | dhi ~ mhi;
box {rot} (mlo, mmid, mhi) | dlo;
cnot dhi | mhi;
cnot dhi | dlo ~ mhi;
output {$s_r^\text{hi}$} dhi;
output {$s_r^\text{mid}$} dlo;
output {$s_r^\text{lo}$} mhi;
output {$i^\text{hi}$} mmid;
output {$i^\text{lo}$} mlo;
\end{yquant}
\end{tikzpicture}.
\end{equation}
The first $X$ together with the two cnots ensure that $d^\text{lo}$ is set to zero for all entries on the diagonal. As indicated in the circuit, that qubit is then $\ket{0}$ for all valid inputs $(d,m)$. It is then used as a control qubit for a rotation implementing $\lfloor m/2 \rfloor$ from \eqref{eq:i tree}. The rotation consists of swaps circularly rotating the qubits down. Finally, the cnot and ccnot ensure that $0\le s_r < S_r=4$, i.e.~$s_r^\text{hi}=0$ for all valid inputs. Specifically, the enumeration by $s_r$ within rows implemented by the circuit is
\begin{equation}
\begin{pmatrix}
0 & 3 & & & & & & \\
1 & 0 & 2 & 3 & & & & \\
    & 1 & 0 & & 2 & 3 & & \\
    & 1 & & 0 & & & 2 & 3 \\
    &  & 1 & & 0 & & & \\
    &  & 1 &  & & 0 & & \\
    &  &  & 1 & & & 0 & \\
    &  &  & 1 & & &  & 0
\end{pmatrix}.
\end{equation}
The out-of-range oracle $O_\text{rg}$ is:
\begin{equation}
\begin{tikzpicture}
\begin{yquant}
qubit del;
qubit {$d^\text{hi}$} dhi;
qubit {$d^\text{lo}$} dlo;
qubit {$m^\text{hi}$} mhi;
qubit {$m^\text{mid}$} mmid;
qubit {$m^\text{lo}$} mlo;
["north:$N/2$" {font=\protect\footnotesize, inner sep=0pt}]
slash mlo;
[plusctrl, shape=yquant-circle]
box {$O_\text{rg}$} (dhi,dlo,mhi,mmid,mlo) | del;

text {\ =\ } (-);
init {del} del;
init {$d^\text{hi}$} dhi;
init {$d^\text{lo}$} dlo;
init {$m^\text{hi}$} mhi;
init {$m^\text{mid}$} mmid;
init {$m^\text{lo}$} mlo;

cnot del | mhi ~ dhi, dlo;
cnot del  ~ mhi, mmid, dhi, dlo;
cnot del  ~ mhi, mmid, mlo, dhi, dlo;
cnot del | dlo ~ dhi, mhi, mmid, mlo;
cnot del | dlo, mmid ~ dhi;
cnot del | dlo, mhi ~ dhi, mmid;
cnot del | dhi ~ dlo, mhi, mlo;
cnot del | dhi,dlo;
\end{yquant}
\end{tikzpicture}
\end{equation}

From these oracles, a Hermitian base and preamplified block encoding can be constructed with $S_r=4$. 

One cannot use the PREP/UNPREP scheme with above oracles as the multiplexed rotations do not commute with $O_r$. This has to do with the fact that in \eqref{eq:circuit prep block encoding}, $H_{S_r/D} = H_{4/3}$ does not exist. One can artifically increase $S_r$ to 6 and then construct a PREP/UNPREP block encoding circuit. It would have subnormalisation $2(|A_0|+|A_1|+|A_2|)$.

\subsection{2-dimensional Laplacian}
Here, we construct an example block-encoding of a 2-dimensional discrete Laplacian operator on a grid. Using finite differences, the one-dimensional Laplacian operator can be expressed as
\begin{equation}
    \Delta f(x_i) = \frac{f(x_{i-1}) - 2 f(x_i) + f(x_{i+1})}{(\Delta x)^2}.
\end{equation}
When the discrete values $f(x_i)$ are understood as the components of a vector, the Laplacian corresponds to a Toeplitz matrix with $-2/(\Delta x)^2$ on the diagonal, and $+1/(\Delta x)^2$ on the first two offdiagonals. The block encoding scheme in \ref{sec:toeplitz} can be used. Now, let us consider a two-dimensional regular grid of size $N_x\times N_y$, with dimensions powers of 2 for simplicity as usual. The finite difference Laplacian becomes
\begin{equation}
    \Delta f(x_a, y_b) = \frac{f(x_{a-1},y_b) - 2 f(x_a, y_b) + f(x_{a+1}, y_b)}{(\Delta x)^2} + \frac{f(x_a,y_{b-1}) - 2 f(x_a, y_b) + f(x_{a}, y_{b+1})}{(\Delta y)^2}.
\end{equation}
A standard encoding of the values on the grid into an $N=N_xN_y$-dimensional vector is row by row:
\begin{equation}
    f_{a+bN_x} = f(x_a,y_b),\ a\in{0,\ldots, N_x-1},\ b\in{0,\ldots,N_y-1}
    \label{eq:2d encoding}
\end{equation}
Then, the Laplacian matrix $A$ is defined by
\begin{equation}
    A_{a_1+b_1N_x, a_2+b_2N_x} = \begin{cases} A_0 := -2(1/(\Delta x)^2 + 1/(\Delta y)^2) & \text{for}\ a_1=a_2,\ b_1=b_2 \\
    A_1 := 1/(\Delta x)^2 & \text{for}\ |a_1-a_2|=1,\  b_1=b_2 \\
    A_2 := 1/(\Delta y)^2  & \text{for}\ |b_1-b_2|=1,\ a_1=a_2 \\
    0 & \text{else}
    \end{cases}
\end{equation}
We have 3 different values, which is padded to $D=4$. Visually, for $N_x=4, N_y=4$, the matrix looks like
\begin{equation}
\begin{tikzpicture}
  \matrix (m)[
    matrix of math nodes,
    nodes in empty cells,
    left delimiter=(,
    right delimiter=)
  ] {
       A_0    & A_1  &  &  &  A_2     & &    \\
       A_1 & A_0   & A_1 & &  & A_2    & &  \\
       & A_1    &  A_0  & A_1  &   &    & A_2  \\
       & & A_1 & A_0 & & & & A_2 \\
       A_2&&&& A_0    & A_1  &&& A_2   \\
       &A_2&&& A_1 & A_0   & A_1 &&& A_2    \\
       &&A_2&& & A_1    &  A_0  & A_1&&&A_2  \\
       &&&A_2& & & A_1 & A_0  &&&& A_2 \\
       &&&&A_2&&&& A_0    & A_1  &  &  &  A_2     & &    \\
       &&&&&A_2&&&A_1 & A_0   & A_1 & &  & A_2    & &  \\
       &&&&&&A_2&&& A_1    &  A_0  & A_1  &   &    & A_2  \\
       &&&&&&&A_2&& & A_1 & A_0 & & & & A_2 \\
       &&&&&&&&A_2&&&& A_0    & A_1     \\
       &&&&&&&&&A_2&&&A_1 & A_0   & A_1  \\
       &&&&&&&&&&A_2&&& A_1    &  A_0  & A_1  \\
       &&&&&&&&&&&A_2&& & A_1 & A_0 \\
  };
  \draw[dashed] (m-1-1.north west) rectangle (m-4-4.south east);
  \draw[dashed] (m-5-5.north west) rectangle (m-8-8.south east);
  \draw[dashed] (m-9-9.north west) rectangle (m-12-12.south east);
  \draw[dashed] (m-13-13.north west) rectangle (m-16-16.south east);
\end{tikzpicture}
\label{eq:poisson structure}
\end{equation}
where there are $N_y$ dashed rectangles of dimension $N_x\times N_x$ each. We have maximum sparsity 5 padded to $S=8$, and maximum multiplicity $2N-2$ padded to $M=2N$, such that $NS=DM=8N$.
For the labelling by $m$, we separate the high bit of $m^\text{hi}$ from the low $\log_2 N$ bits $m^\text{lo}$. We choose $m^\text{hi}=0$ for the lower left triangular matrix (including the diagonal), and $m^\text{hi}=1$ for the upper triangular matrix. We choose $m^\text{lo}$ to be the row index (in the lower triangular matrix) or the column index (in the upper triangular matrix).

The Hermitian block encoding scheme for symmetric matrices (section~\ref{sec:hermitian}) requires a transposition oracle $O_t(d,m)=(d,m')$ that gives the corresponding label of the transposed element. We have
\begin{equation}
    O_t(d,m^\text{hi},m^\text{lo})=\begin{cases} (d,m^\text{hi},m^\text{lo}) & \text{for}\ d=0 \\
    (d, 1-m^\text{hi},m^\text{lo}) & \text{for}\ d=1\ \text{or}\ d=2 \end{cases}
\end{equation}
which in quantum circuit form is
\begin{equation}
\begin{tikzpicture}
\begin{yquant}
qubit {$d^\text{hi}$} dhi;
qubit {$d^\text{lo}$} dlo;
qubit {$m^\text{hi}$} mhi;
qubit {$m^\text{lo}$} mlo;
["north:$N$" {font=\protect\footnotesize, inner sep=0pt}]
slash mlo;
box {$O_t$} (-);
text {$=$} (-);
slash mlo;
cnot mhi | dlo;
cnot mhi | dhi;
\end{yquant}
\end{tikzpicture}
\end{equation}
Next, we need a column oracle $O_c: (d,m) \to j$ that gives the column index.
\begin{equation}
    O_c(d, m^\text{hi}, m^\text{lo}) = \begin{cases}  
    m^\text{lo} & \text{for}\ d=0 \\
    m^\text{lo} & \text{for}\ m^\text{hi} = 1 \\
    m^\text{lo}-1 & \text{for}\ m^\text{hi}=0\ \text{and}\ d=1\\
    m^\text{lo}-N_x & \text{for}\ m^\text{hi}=0\ \text{and}\ d=2 
    \end{cases}
\end{equation}
In quantum circuit form this is
\begin{equation}
\begin{tikzpicture}
\begin{yquant}
qubit {$d^\text{hi}$} dhi;
qubit {$d^\text{lo}$} dlo;
qubit {$m^\text{hi}$} mhi;
qubit {$m^\text{lo}$} mlo;
["north:$N$" {font=\protect\footnotesize, inner sep=0pt}]
slash mlo;
box {$O_c$} (-);
text {$=$} (-);

["north:$N$" {font=\protect\footnotesize, inner sep=0pt}]
slash mlo;
align -;
text {$d^\text{hi}$} dhi;
text {$d^\text{lo}$} dlo;
text {$m^\text{hi}$} mhi;
text {$m^\text{lo}$} mlo;

box {$-1$} mlo | dlo ~ mhi;
box {$-N_x$} mlo | dhi ~ mhi;

text {$s^\text{hi}$} dhi;
text {$s^\text{mid}$} dlo;
text {$s^\text{lo}$} mhi;
text {$j$} mlo;
\end{yquant}
\end{tikzpicture}
\end{equation}
and the $\ket{s}$ states required for $H_{S}$ are $H_{S} = \frac{1}{\sqrt 5}(\ket{000}+\ket{010}+\ket{011}+\ket{100}+\ket{101})$ (reading $s^\text{hi}$ to $s^\text{lo}$ left to right).

The out-of-range indices $(d,m)$ are the following:
\begin{equation}
    (d,m^\text{hi}, m^\text{lo})\ \text{out-of-range for}\ \begin{cases}
        d=0, m^\text{hi}=1 \\
        d=1, (m^\text{lo}\ \text{mod}\ N_x) = 0 \\
        d=2, m^\text{lo} < N_x \\
        d=3
    \end{cases}
\end{equation}
To write this in quantum circuit form, we split the $m^\text{lo}$ register into two registers of $\log_2 N_y$ and $\log_2 N_x$ bits; this respects the structure in \eqref{eq:2d encoding}, and makes it easy to to implement the conditions $(m^\text{lo}\ \text{mod}\ N_x) = 0$ and $m^\text{lo}<N_x$. (Splitting $j$ into those two registers also simplifies the $-N_x$ in $O_c$, which is just a $-1$ on the $\log_2 N_y$-bit register.) The out-of-range oracle is
\begin{equation}
\begin{tikzpicture}
\begin{yquant}
qubit del;
qubit {$d^\text{hi}$} dhi;
qubit {$d^\text{lo}$} dlo;
qubit {$m^\text{hi}$} mhi;
qubit {} my;
qubit {} mx;
init {$m^\text{lo}$} (my, mx);
["north:$N_y$" {font=\protect\footnotesize, inner sep=0pt}]
slash my;
["north:$N_x$" {font=\protect\footnotesize, inner sep=0pt}]
slash mx;

[plusctrl, shape=yquant-circle]
box {$O_\text{rg}$} (dhi, dlo, mhi, my, mx) ~ del;

hspace {3pt} -;
discard -;

text {$=$} (-);

init {del} del;
init {$d^\text{hi}$} dhi;
init {$d^\text{lo}$} dlo;
init {$m^\text{hi}$} mhi;
init {$m^\text{lo}$} (my, mx);
["north:$N_y$" {font=\protect\footnotesize, inner sep=0pt}]
slash my;
["north:$N_x$" {font=\protect\footnotesize, inner sep=0pt}]
slash mx;
cnot del | mhi ~ dlo, dhi;
cnot del | dlo ~ dhi, mx;
cnot del | dhi ~ dlo, my;
cnot del | dhi, dlo;

hspace {3pt} -;
discard -;

text {$\hat=$} (-);

init {del} del;
init {$d^\text{hi}$} dhi;
init {$d^\text{lo}$} dlo;
init {$m^\text{hi}$} mhi;
init {$m^\text{lo}$} (my, mx);
["north:$N_y$" {font=\protect\footnotesize, inner sep=0pt}]
slash my;
["north:$N_x$" {font=\protect\footnotesize, inner sep=0pt}]
slash mx;
cnot del | mhi ~ dlo, dhi;
cnot del | dlo ~ mx;
cnot del | dhi ~ my;
\end{yquant}
\end{tikzpicture}
\end{equation}
When putting together the circuit, we can use the simplified form on the right because certain controls are never triggered.

The full Hermitian block encoding circuit \eqref{eq:hermitian block encoding circuit} for these oracles is then:
\begin{equation}
\begin{tikzpicture}
\begin{yquant}
qubit data;
qubit del;
qubit {$s^\text{hi}$} dhi;
qubit {$s^\text{mid}$} dlo;
qubit {$s^\text{lo}$} mhi;
qubit {} my;
qubit {} mx;
init {$j$} (my, mx);
["north:$N_y$" {font=\protect\footnotesize, inner sep=0pt}]
slash my;
["north:$N_x$" {font=\protect\footnotesize, inner sep=0pt}]
slash mx;

box {$H_S$} (dhi, dlo, mhi);

box {$+1$} my | dhi ~ mhi;
box {$+1$} (my, mx) | dlo ~ mhi;

align -;
init {$d^\text{hi}$} dhi;
init {$d^\text{lo}$} dlo;
init {$m^\text{hi}$} mhi;
init {$m^\text{lo}$} (my, mx);

align -;

z data;

cnot del | mhi ~ dlo, dhi;
cnot del | dlo ~ mx;
cnot del | dhi ~ my;

[multictrl]
box {$R(A_d/||A||_\text{max}$} data ~ dhi, dlo;
align -;

cnot mhi | dlo;
cnot mhi | dhi;

box {$-1$} (my, mx) | dlo ~ mhi;
box {$-1$} my | dhi ~ mhi;

box {$H_S^\dag$} (dhi, dlo, mhi);

output {$i$} (my, mx);
\end{yquant}
\end{tikzpicture}
\label{eq:2d laplace circuit}
\end{equation}
The subnormalisation is 
\begin{equation}
    \alpha = 5 \max(|A_0|,|A_1|,|A_2|) = 5|A_0| = 10\left(\frac{1}{(\Delta x)^2} + \frac{1}{(\Delta y)^2}\right).
\end{equation}
Since the data loading and row/column oracles commute, we can use the PREP/UNPREP scheme to improve the subnormalisation to (in the symmetric $p=1/2$ case)
\begin{equation}
    |A_0| + |A_1| +|A_2| = 3\left(\frac{1}{(\Delta x)^2} + \frac{1}{(\Delta y)^2}\right).
\end{equation}

Block encodings of a 2-dimensional Laplacian for use in quantum algorithms were already considered in \cite{camps_fable_2022}, using an approach that constructs an approximate block encoding. Here, in contrast, the block encoding is exact (up to finite accuracy in the data loading oracle). Moreover, the block encoding constructed with our scheme requires $O(\log N+\log N_y)$ gates, which comes from the additions in \eqref{eq:2d laplace circuit}. Whereas, the approximate block encoding in \cite{camps_fable_2022} appears to have an exponentially worse scaling. Our method does well, because treating it solely as a sparse \cite{gilyen_quantum_2019} or dense matrix \cite{clader_quantum_2022,chakraborty_power_2019} does not harness the repeated elements, and it is not of the class considered in \cite{camps_explicit_2022}.

\section{Conclusions and Outlook}
\label{sec:conclusions}

In this work, we have presented a number of schemes to block encode structured matrices (base, preamplified, PREP/UNPREP schemes, and Hermitian extensions). Such a block encoding is necessary to use a matrix in QSVT and related quantum algorithms. All the schemes are based on a labelling of the matrix structure in terms of $(d,m)$, where $d$ labels distinct non-zero values and $m$ distinguishes different elements with the same value. Arithmetic quantum circuits relating this labelling to the column and row indices constitute the core of the quantum circuits. Section~\ref{sec:block encodings} shows our circuit constructions based on these arithmetic oracles along with a data loading oracle. 

All schemes fully incorporate the sparsity of the matrix, reflected in the flag qubit number, as well as repeated values: The data loading oracle is controlled on $d$, such that the data loading cost corresponds to the number of distinct values; no value is loaded twice, even if it appears in the structured matrix multiple times. The schemes differ in the subnormalisation achieved; our detailed analysis is summarised in table~\ref{table:summary}. Which scheme performs best depends on the specific matrix in question. Further, our block encodings can be adapted to be Hermitian in the case of symmetric matrices (section~\ref{sec:hermitian}) without extra data loading cost or subnormalisation.

In section~\ref{sec:examples} we have provided examples showing our block-encoded schemes in action for several families of structured matrices (Toeplitz matrix, tridiagonal matrix, extended binary tree matrix, 2d Laplacian matrix).  Together with the information provided in the appendix on data loading, the full circuits can be elaborated. We hope that, beyond the examples considered here, our schemes will prove useful for constructing block encodings for a large variety of matrix families from different application areas.

A theoretical bound for the subnormalisation is the spectral norm $||A||_{op}$, due to the requirement that the block encoding be unitary. None of the schemes achieve this for an arbitrary matrix, so improvements may still be possible.
While this article focuses on real matrices, an extension to complex valued matrices is straightforward: One could sum block encodings of a matrices real and imaginary parts, or adapt the multiplexed rotations to yield complex amplitudes of the $\ket{0}$ state.
We have assumed a structure in the pattern and position of matrix elements, but not in the values. If the matrix possesses further structure, we expect more efficient circuits can be found. For example, if the values depend arithmetically on any of the indices $i,j,d,m$, data loading may not be necessary. Further, the block encodings constructed are exact, up to finite accuracy in the data loading (see appendix~\ref{sec:data loading}) and, in the preamplification scheme, the accuracy of singular value amplification (see appendix~\ref{sec:singular value amplification}). Possibly, approximate block encodings \cite{camps_fable_2022,nguyen_block-encoding_2022} could be implemented with more efficient circuits.

\section*{Acknowledgements}
We thank Leigh Lapworth of Rolls-Royce plc for inspiration of matrix structures. Work partially
funded by the UK’s Commercialising Quantum Technologies Programme (Grant reference 10004857).

\bibliography{main}

\appendix

\section{Implementation of data loading step}
\label{sec:data loading}

An important component in all of the block encodings is data loading, in which the values of matrix elements are loaded into the matrix structure supplied by the other, arithmetic, oracles. In fact, we use the
data loading cost (number of data items) as a stand-in for the complexity of a block encoding circuit. 
In this appendix, we will give details on the data loading step.

\subsection{Multiplexed rotations}
\label{sec:multiplexed rotations}

In the base and preamplified schemes (sections~\ref{sec:base block encoding} and \ref{sec:preamplified}), data loading is accomplished with a data loading oracle $O_\text{data}$ implementing multiplexed rotations \eqref{eq:formula multiplexed rotations}:
\begin{equation}
    O_\text{data} = \sum_{d=0}^{D-1}
    R_X(2\arccos A_d / ||A||_\text{max})\otimes\ket{d}\!\bra{d}.
    \label{eq:formula multiplexed rotations appendix}
\end{equation}
It has been shown \cite{mottonen_quantum_2004,shende_synthesis_2006} how such rotations multiplexed on $D$ values can be decomposed into $D$ uncontrolled rotation and $D$ CNOT gates (for $D$ a power of 2), without use of any ancillas. The $T$ gate cost is then that of synthesizing $D$ arbitrary rotations into Clifford and $T$ gates up to the required accuracy $ε$. Such synthesis requires $O(\log (1/ε))$ $T$ gates \cite{ross_optimal_2016}, such that the total $T$ gate count is the product
\begin{equation}
    D\cdot O(\log (1/ε)). \label{eq:cost multiplexed rotations}
\end{equation}

The multiplicative factor $O(\log(1/ε))$ can be changed to additive, reducing overall $T$ count. To this end, the (approximate) bitvalues of all of the required angles are first loaded into ancilla qubits, and rotations then performed according to the loaded bitvalues.
Accordingly, this implementation of $O_\text{data}$ consists of 3 steps \cite{von_burg_quantum_2021,low_halving_2021,low_trading_2018}:
\begin{align}
    \ket{0}\ket{d}\ket{0}^{\otimes b} \xrightarrow{(1)} \ket{0}\ket{d}\ket{α_d} & \xrightarrow{(2)} (R_X(α_d)\ket{0})\ket{d}\ket{α_d} 
    \xrightarrow{(3)} (R_X(α_d)\ket{0})\ket{d}\ket{0}^{\otimes b},
\end{align}
with $α_d = 2 \arccos A_d/||A||_\text{max}$.

Step (1) loads a binary representation of $α_d$ into the $b$-qubit ancilla register ($b\sim\log_2 1/ε$). Using additional $\lceil\log_2 D\rceil$ ancilla qubits, this can be done with $D-2$ Toffoli gates with the QROM presented in \cite{babbush_encoding_2018}. Note that for QROM, $D$ needn't be a power of 2. If further $\sim\sqrt{D}$ ancilla qubits are available, the select-swap technique \cite{low_trading_2018} (also dubbed QROAM \cite{berry_qubitization_2019}) can be used to lower Toffoli count to $\sim\sqrt{D}$.

Next, step (2) rotates the data qubit according to the bitvalue saved in the ancilla register. This can be done with the phase-gradient technique \cite{von_burg_quantum_2021,low_halving_2021,low_trading_2018}, in which the $b$-bit angle register is added into a reusable phase gradient state. This addition requires $b-1$ Toffolis. Alternatively, step (2) can be implemented with $b$ rotations controlled on each of the qubits in $\ket{α_d}$ in turn \cite{low_trading_2018}.

Finally, in step (3), the data lookup must be uncomputed to return the ancillas to $\ket{0}^{\otimes b}$. In some cases, the Toffoli cost of uncomputing the lookup can be reduced compared to step (1) by using measurement based uncomputation \cite{babbush_encoding_2018}.

The cost of the three-step procedure is asymptotically reduced from \eqref{eq:cost multiplexed rotations} because steps (1), (2), and (3) are in sequence and the term related to the accuracy of the rotations is added rather than multiplied to the term $D$ related to the data look-up. In either case, the $T$ cost scales with $D$ as $O(D)$ when using QROM; the number of data items to load is a sensible stand-in for circuit length ($T$ count) while staying agnostic about the exact procedure.

\subsection{State preparation}
\label{sec:state preparation}

In the PREP/UNPREP scheme (section~\ref{sec:prep unprep block encoding}), the data loading oracle $O_\text{data}$ and diffusion operator $H_S$ are merged into a state preparation operator, reducing the subnormalisation. The data loading is not performed with multiplexed rotations $O_\text{data}$ on a data flag qubit like in the base and preamplified scheme. Instead, data is loaded as the amplitudes of a state that is prepared with a PREP operator.

An arbitrary quantum state $\ket{ψ} = \text{PREP}\ket{0}^{\otimes\log_2 D}$ can be implemented by a sequence of multiplexed rotations, whose angles $α$ have been precomputed from $ψ$'s amplitudes \cite{shende_synthesis_2006,low_trading_2018}, like in this example with $D=8$:
\begin{equation}
\begin{tikzpicture}
\begin{yquant}
qubit {$\ket{0}$} s[3];
box {$R_Y(α)$} s[0];
[multictrl]
box {$R_Y(α_\cdot)$} s[1] ~ s[0];
[multictrl]
box {$R_Y(α_\cdot)$} s[2] ~ s[0], s[1];
output {$\ket{\psi}$} (-);
\end{yquant}
\end{tikzpicture}
\end{equation}
These multiplexed rotations can be implemented as in appendix~\ref{sec:multiplexed rotations}.
Alternative approaches to preparing $\ket{\psi}$ include coherent alias sampling \cite{babbush_encoding_2018} and prerotation \cite{clader_quantum_2022}.

\subsection{Data loading oracle in Gily\'{e}n et.~al.'s sparse scheme}\label{sec:sparse_data_loading}

Lemma 48 in \cite{gilyen_quantum_2019} shows a construction of a block encoding in terms of black-box oracles. One of the black boxes is the data loading oracle called $O_A$. Given the row and column indices $i$ and $j$, it loads the $b$-bit bitstring $A_{ij}$ of the corresponding matrix entry:
\begin{equation}
O_A\ket{i}\ket{j}\ket{0}^{\otimes b} = \ket{i}\ket{j}\ket{A_{ij}}.
\end{equation}
The implementation of this oracle is not discussed.

From a first cursory look at the oracle, one might conclude that repeated entries must be loaded separately. However, given a structured matrix as considered in this paper, it can be implemented with only $D$ data loading and the usual $O(\text{polylog}\ N)$ arithmetic overheads:
\begin{equation}
\begin{tikzpicture}
\begin{yquant}
qubit {$\ket{0}^{\otimes b}$} data;
qubit {$\ket{i}$} i;
qubit {$\ket{j}$} j;
qubit {$\ket{0}$} d;
qubit {$\ket{0}$} m;
["north:$N$" {font=\protect\footnotesize, inner sep=0pt}]
slash i, j;
["north:$D$" {font=\protect\footnotesize, inner sep=0pt}]
slash d;
["north:$M$" {font=\protect\footnotesize, inner sep=0pt}]
slash m;

box {$O_{rc}$} (i,j,d,m);
text {$i$} i;
text {$j$} j;
text {$d$} d;
text {$m$} m;

[multictrl]
box {$\ket{0}^{\otimes b} \mapsto \ket{A_{d}=A_{ij}}$} data ~ d;

box {$O_{rc}^\dag$} (i,j,d,m);

output {$\ket{i}$} i;
output {$\ket{j}$} j;
output {$\ket{0}$} d,m;
\end{yquant}
\end{tikzpicture}
\end{equation}
The arithmetic oracle $O_{rc}$ computes $(d,m)$ into ancilla qubits from $(i,j)$. Possibly, some of the arithmetics could be done in-place, and the $i$ and $j$ registers reused for $d$ and $m$.

\section{Singular value amplification}
\label{sec:singular value amplification}

Singular value amplification \cite{gilyen_quantum_2019,low_hamiltonian_2017} allows to reduce the subnormalisation of a block encoding by increasing the circuit length. This is achieved by performing a QSVT that multiplies the singular values by an amplification factor $γ$. While the resulting block encoding (with one flag qubit more) will have a subnormalisation reduced by a factor of $γ$, the circuit length will have roughly increased by a factor of $γ$, up to logarithmic factors and constants.
This reciprocal behaviour substantiates the figure of merit 
\begin{equation}
    \text{circuit cost} \cdot \text{subnormalisation}\,,
\end{equation}
of which eq.~\eqref{eq:figure of merit} is a proxy, used to assess a block encoding.

As a crucial ingredient to the preamplified scheme (section~\ref{sec:preamplified}), the complexity of singular value amplification including constant factors determines which block encoding scheme performs best for specific data. While previous work has determined its big-$O$ scaling \cite{gilyen_quantum_2019,low_hamiltonian_2017}, this appendix will shed light on the constant factors.

\begin{theorem}[Uniform singular value amplification (adapted from Theorem 20, \cite{gilyen_quantum_2019})]
Let $U$ be a block encoding, $γ>1$, and $0<δ,ε<1/2$, whose block-encoded matrix $A$ has singular values $ζ_i\in[0, (1-δ)/γ]$. Then there is an efficiently computable polynomial of degree
\begin{equation}
    O\left(\frac{γ}{δ}\log\frac{γ}{ε}\right)
    \label{eq:asymptotic degree}
\end{equation}
whose application to $U$ by QSVT results in a block encoding $\tilde U$ encoding $\tilde A$, an $ε$-approximation of $γ A$.
In particular, each amplified singular value $\tilde\zeta_i$ of $\tilde A$ has relative accuracy $ε$:
$|\tilde\zeta_i / (\gamma ζ_i) - 1| \le ε$.
\end{theorem}
Note that singular value amplification can be applied to general projected matrices beyond quadratic $A$ flagged by $\ket{0}$s. As such, it encompasses uniform amplitude amplification.

\begin{figure}
    \centering
    \includegraphics[width=13cm]{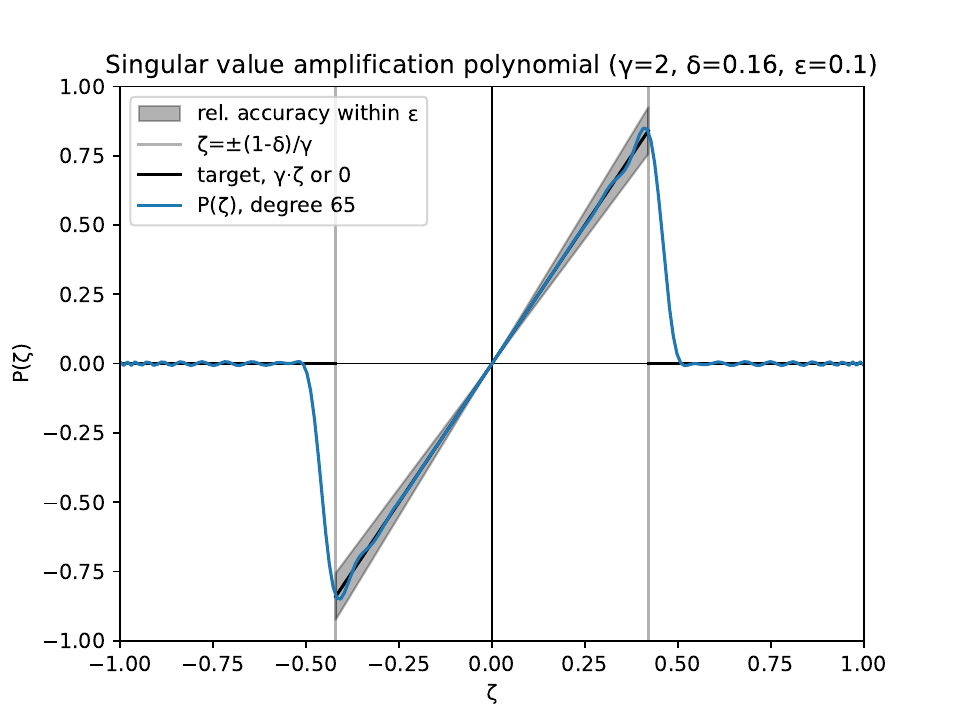}
    \caption{Singular value amplification. The target function for an amplification by $γ$ is $γ\zeta$ (black line), and 0 outside of the validity region. For QSVT, it has to be approximated by a polynomial (blue) of sufficient accuracy. The region of accuracy (grey shaded) is determined by the parameters $δ$ and $ε$. Asymptotically, the degree was shown to be $O(γ/\delta\,\log (γ/ε))$ \cite{low_hamiltonian_2017}. We truncate a Chebyshev expansion of an analytic approximation to the target function to find the polynomial's degree.} 
    \label{fig:singular value amplification}
\end{figure}

Figure~\ref{fig:singular value amplification} shows an example of a target function for amplification, along with a polynomial approximation for the QSVT.
The parameter $ε$ controls the relative accuracy of the singular value amplification, while $δ$ controls the applicable range. The requirement $\zeta_i \le 1/γ$ is natural because the singular values $\tilde\zeta_i\approx γ\zeta_i$ of the block-encoding $\tilde U$ must be bounded by one. To ensure the polynomial approximation can be bounded by one across the entire range $[-1,+1]$ (a requirement for QSVT), the permissible range of singular values of $A$ must be slightly lowered by $δ$. In principle, a trade-off between $γ$ and $δ$ is possible. For preamplification, Gily\'{e}n et.~al.~\cite{gilyen_quantum_2019} choose an amplification factor of
\begin{equation}
    γ = \frac{1}{\sqrt[4]{2}}\frac{1}{\zeta_\text{max}} \approx 0.84 \frac{1}{\zeta_\text{max}}\ (\text{with}\ \zeta_\text{max} = \max_i \zeta_i),
\end{equation}
which leads to
\begin{equation}
    δ = 1 - γ\zeta_\text{max} = 1 - \frac{1}{\sqrt[4]{2}} \approx 0.16. \label{eq:min delta}
\end{equation}

The starting point for the polynomial $P(\zeta)$ is a sufficiently good analytic approximation of the rectangle function on the domain $[-(1-δ)/γ,(1-δ)/γ]$ based on error functions, see \cite{gilyen_quantum_2019,low_hamiltonian_2017}. Its expansion in Chebyshev polynomials is truncated such that an absolute accuracy of $ε$ is achieved. The product with $γ\zeta$ then gives the desired polynomial approximation. Yet, the authors of \cite{low_hamiltonian_2017}
``emphasize that our proposed sequence of polynomial transformations serve primarily to prove their asymptotic scaling.'' They suggest to obtain the constant
factors by a direct Chebyshev truncation of the entire functions.

\begin{figure} 
    \centering
    \includegraphics[width=7cm]{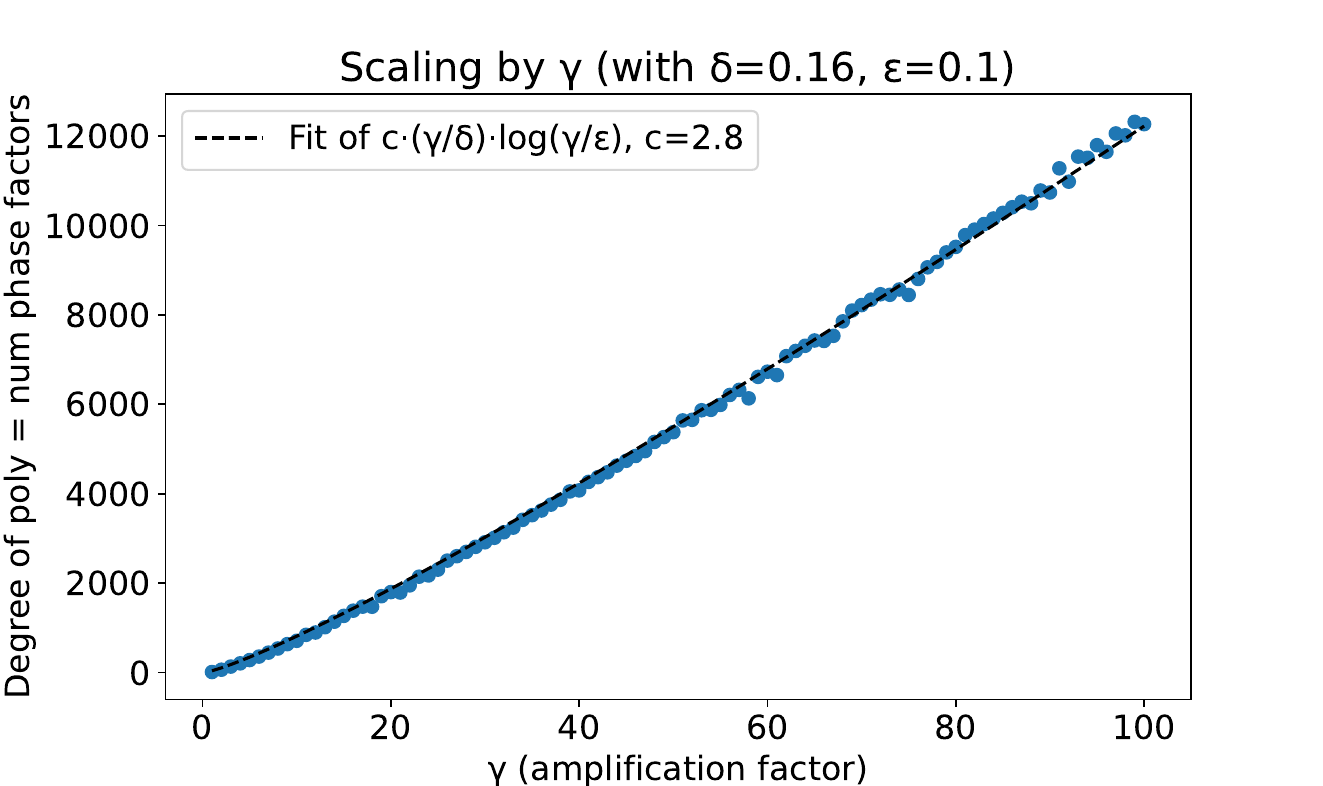}
    \includegraphics[width=7cm]{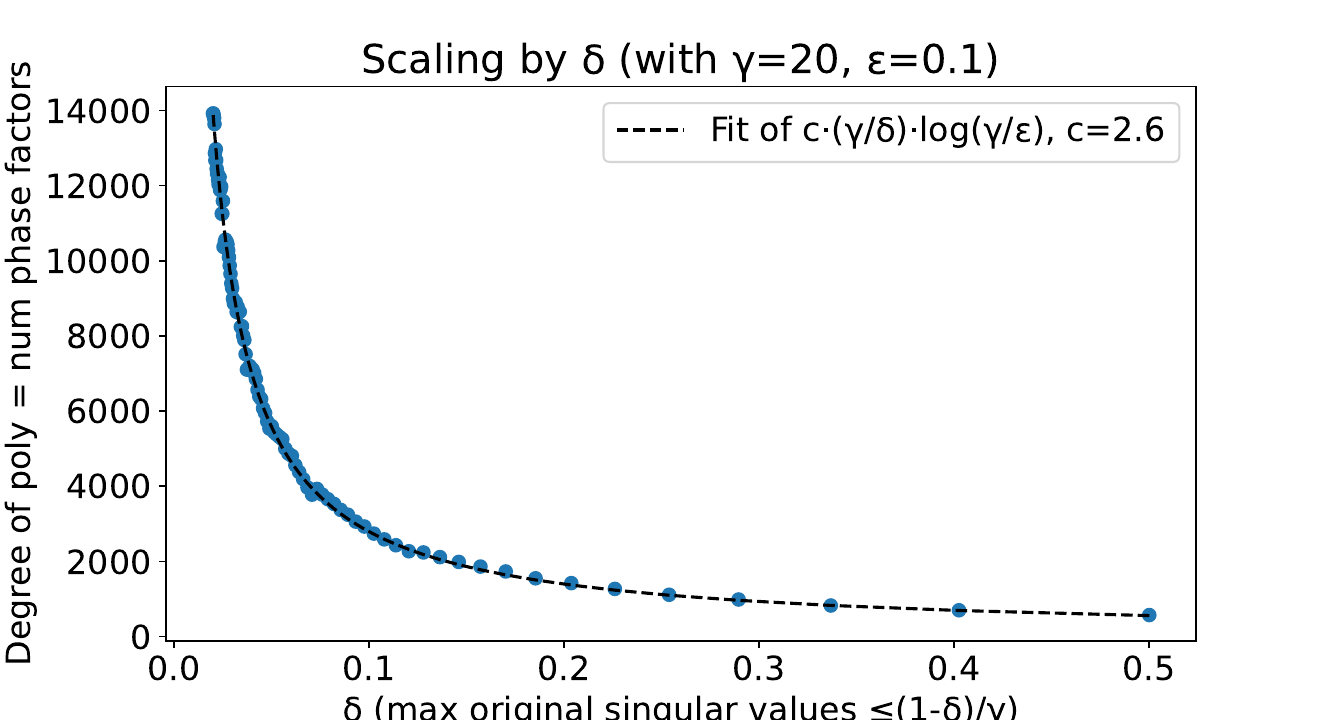}
    \includegraphics[width=7cm]{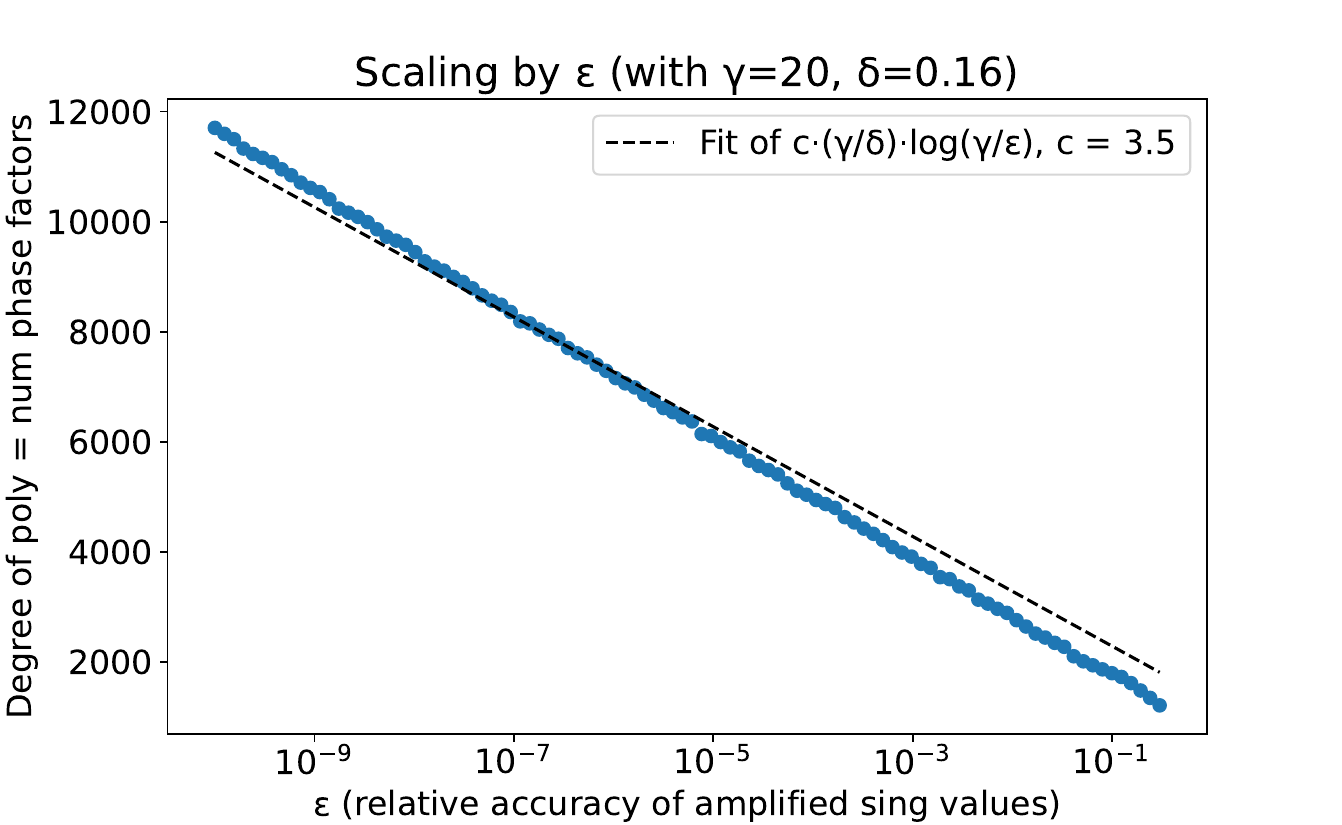} 
    \caption{Degree of the polynomial for singular value amplification. Asymptotically, the degree was shown to be $O(γ/\delta\,\log (γ/ε))$ \cite{low_hamiltonian_2017}. We truncate a Chebyshev expansion of an analytic approximation to the target function to find the polynomials' exact degrees. By varying the amplification factor $γ$, the accuracy $ε$, and $δ$, we aim to find the constant factor in the asymptotic complexity. From the fits shown in the plots, we conclude the degree is approximately $3\,γ/δ\,\log(γ/ε)$. Log refers to the natural logarithm.}
    \label{fig:degree scaling}
\end{figure}

Here, we therefore numerically perform Chebyshev truncations of the entire approximation to the rectangle function \footnote{The Chebyshev truncation was performed with numpy's \texttt{np.polynomial.Chebyshev.interpolate()} function, and the accuracy of the truncation checked on a grid of $10^3γ/(1-δ)$ points spanning $[-1,+1]$. Calculations were carried out with the default 64bit floating point accuracy.}. The optimal degree satisfying the required accuracy is found with a binary search for various parameter combinations of the amplification factor $γ$, the accuracy $ε$, and $δ$. Fig.~\ref{fig:degree scaling} shows our results along with a fit to the big-$O$ result \eqref{eq:asymptotic degree}. This study confirms the scaling behaviour \eqref{eq:asymptotic degree} and determines a surprisingly small scaling factor of $\approx 3$.
We thus find singular value amplification requires approximately
\begin{equation}
    3\frac{γ}{δ}\log\frac{γ}{ε}
\end{equation}
repetitions of the block encoding, where log is a natural logarithm.

Note that this analysis follows the prescription from \cite{low_hamiltonian_2017}, where the polynomial is constructed from an approximate rectangle function multiplied by $γ\zeta$. Actually, the behaviour of the polynomial outside of the accuracy region does not matter, as long as it stays bounded by $\pm1$. In particular, in Fig.~\ref{fig:singular value amplification} it does not need to decay to zero rapidly around $\zeta=\pm(1-δ)/γ$. Perhaps a lower degree polynomial exists that does not follow the analytic construction based on the rectangle function. Improved degrees to the analytic construction have already been achieved for other target functions (like for matrix inversion) in \cite{dong_efficient_2021}. The authors used a numerical optimisation based approach based on the Remez method to directly find Chebyshev approximations.

\end{document}